\documentclass[
 aps,
 prl,
 amsmath,amssymb,
 reprint,%
footinbib
]{revtex4-2}

\pdfoutput=1
\usepackage{graphicx}
\usepackage{dcolumn}
\usepackage{bm}
\usepackage{empheq}
\usepackage[outdir=./]{epstopdf}
\usepackage{verbatim}
\usepackage{amsmath}
\usepackage{mathtools}
\usepackage{pdfpages}
\usepackage{pgffor}
\newcommand\numberthis{\addtocounter{equation}{1}\tag{\theequation}}

\makeatletter
\newcommand{\Spvek}[2][r]{%
  \gdef\@VORNE{1}
  \left(\hskip-\arraycolsep%
    \begin{array}{#1}\vekSp@lten{#2}\end{array}%
  \hskip-\arraycolsep\right)}

\def\vekSp@lten#1{\xvekSp@lten#1;vekL@stLine;}
\def\vekL@stLine{vekL@stLine}
\def\xvekSp@lten#1;{\def\temp{#1}%
  \ifx\temp\vekL@stLine
  \else
    \ifnum\@VORNE=1\gdef\@VORNE{0}
    \else\@arraycr\fi%
    #1%
    \expandafter\xvekSp@lten
  \fi}
\makeatother

\newcommand{\bq}{\mathbf{q}}

\makeatletter
\AtBeginDocument{\let\LS@rot\@undefined}
\makeatother

\def\supplementfilename{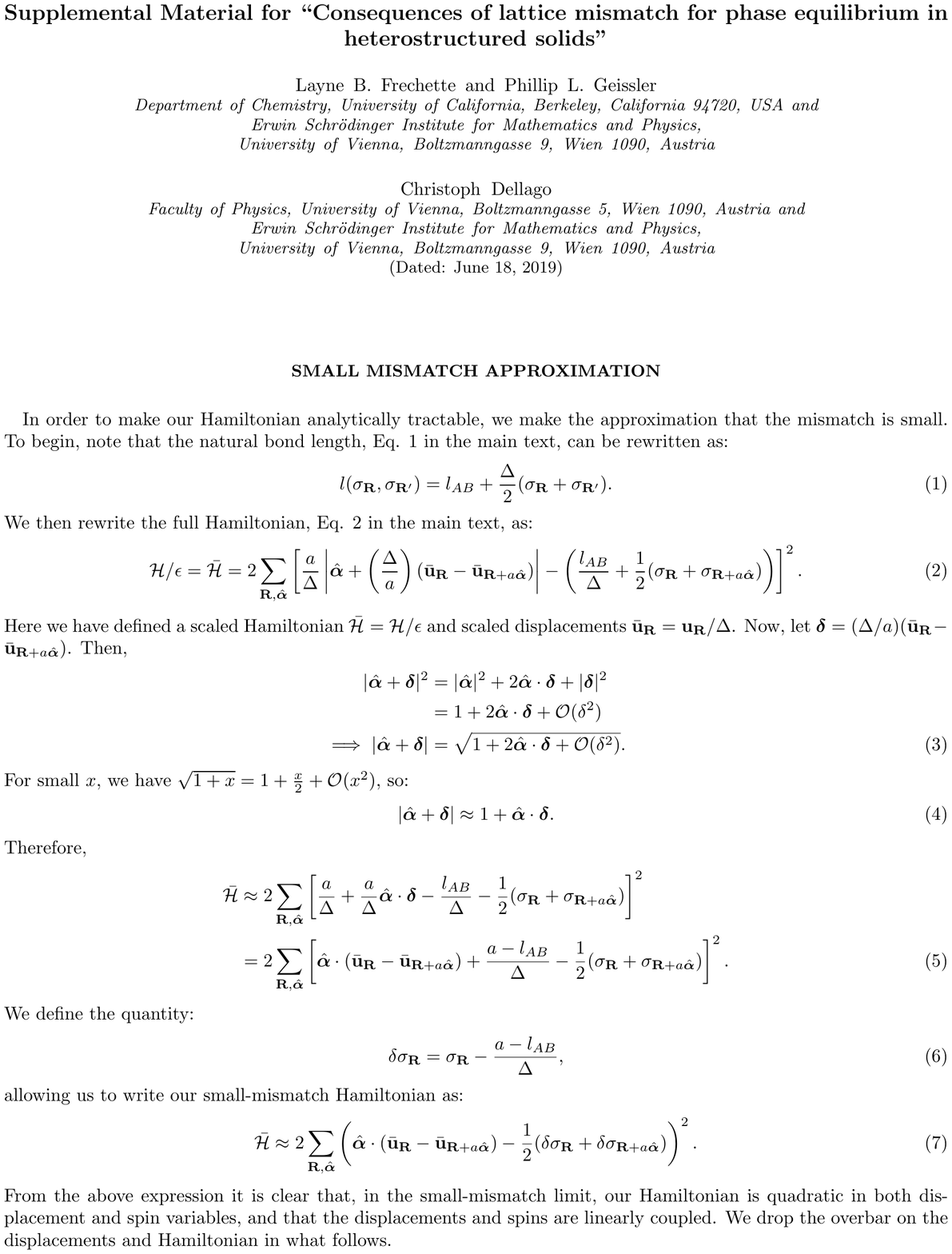}

\pdfximage{\supplementfilename}
\def\numbersupplementpages{\the\pdflastximagepages}

\newif\ifarXiv
\arXivtrue 

\begin{document}

\preprint{AIP/123-QED}

\title{Consequences of lattice mismatch for phase equilibrium in heterostructured solids}

\author{Layne B. Frechette}
 \affiliation{Department of Chemistry, University of California, Berkeley, California 94720, USA}
 \affiliation{Erwin Schr{\"o}dinger Institute for Mathematics and Physics, University of Vienna, Boltzmanngasse 9, Wien 1090, Austria}
\author{Christoph Dellago}%
\affiliation{Faculty of Physics, University of Vienna, Boltzmanngasse 5, Wien 1090, Austria}
\affiliation{Erwin Schr{\"o}dinger Institute for Mathematics and Physics, University of Vienna, Boltzmanngasse 9, Wien 1090, Austria}
\author{Phillip L. Geissler}
 \affiliation{Department of Chemistry, University of California, Berkeley, California 94720, USA}
 \affiliation{Erwin Schr{\"o}dinger Institute for Mathematics and Physics, University of Vienna, Boltzmanngasse 9, Wien 1090, Austria}

\date{\today}

\begin{abstract}
  Lattice mismatch can substantially impact the spatial organization of
  heterogeneous materials. We examine a simple model for
  lattice-mismatched solids over a broad range of temperature and
  composition, revealing both uniform and spatially modulated phases.
  Scenarios for coexistence among them are unconventional due to the extensive
  mechanical cost of segregation.  Together with an adapted Maxwell
  construction for elastic phase separation, mean field theory
  predicts a phase diagram that captures key low-temperature features of Monte Carlo simulations.
\end{abstract}

\maketitle

Lattice mismatch -- the difference in preferred bond length between
adjoining regions of a heterogeneous solid -- is a natural consequence
of mixing diverse components to build complex materials. It is well
recognized that juxtaposing domains with different lattice spacings
introduces local strain, significantly impacting material properties
such as electronic structure \cite{Smith2009,Choi2010,Pereira2009} and
the propensity to form dislocations \cite{Matthews1974,
  People1985}. The resulting elastic energy can also significantly
bias the spatial arrangement of compositional defects and
interfaces. How these biases influence the thermodynamic stability of
mixed phases, however, has not been thoroughly characterized.  Here,
we examine the phase behavior of a microscopic model for
such systems, motivated by intriguing heterostructures adopted by
CdS/Ag$_2$S nanocrystals \cite{Robinson2007} in the course of cation
exchange reactions \cite{Son2004, Li2011, Rivest2013, DeTrizio2016}.  Their
alternating stripes of Cd-rich and Ag-rich domains have been
attributed to lattice mismatch between the CdS and Ag$_2$S domains
\cite{Demchenko2008}, but an understanding of how they form, and
whether they are thermodynamically stable, has been lacking.

Our model and analysis draw from those introduced by Fratzl and
Penrose \cite{Fratzl1995, Fratzl1996}, who represented a two-component
solid by atoms on a flexible square lattice with bond length
preferences that depend on local composition. By integrating out
mechanical fluctuations, they obtained an approximate effective
Hamiltonian for the composition field, whose atomic identities
interact in a pairwise and anisotropic fashion. For the special case
of a 1:1 mixture of the two species, they used mean field theory (MFT) to
predict a second-order phase transition between a high-temperature
disordered phase and a low-temperature ordered phase characterized by
stripes of alternating composition.

This Letter surveys the composition-temperature phase diagram of a
similar model much more broadly, revealing an unanticipated richness
with interesting implications for nanoscale transformations. Monte
Carlo (MC) simulations confirm the predicted appearance of
modulated-order phases with spontaneously broken symmetry. They
further point to unusual scenarios of phase separation, with
well-defined interfaces but a non-convex free energy. This behavior
can be understood as a consequence of elastic energies for phase
separation that scale extensively with system size. For this situation
we devise a procedure, akin to the conventional Maxwell construction,
to determine the boundaries of coexistence regions given equations of
state for the corresponding bulk phases. Although the high temperature phase behavior is dominated by fluctuations on the triangular lattice, a straightforward mean field
theory describes the required bulk properties quite faithfully at low temperature. We
combine these approaches to predict a phase diagram that accounts for
the full set of structures observed in our MC simulations, including
those with system-spanning interfaces.

We consider a model in which $N$ atoms are situated near the sites of
a completely occupied two-dimensional triangular lattice, with
periodic boundary conditions in both Cartesian directions. The atom at
site $\mathbf{R}$ has two possible types, indicated
$\sigma_{\mathbf{R}}=+ 1$ (type $A$) and $\sigma_{\mathbf{R}}= - 1$
(type $B$).  These atom types are distinguished by their size, so that
nearest neighbor atoms at sites $\mathbf{R}$ and
$\mathbf{R}+a\bm{\hat{\alpha}}$ prefer a bond distance $l$ dictated by
their identities,
\begin{equation}
 l(\sigma_{\mathbf{R}},\sigma_{\mathbf{R}+a\bm{\hat{\alpha}}}) = \begin{cases} l_{AA},\,\,\text{for}\,\, \sigma_{\mathbf{R}}=\sigma_{\mathbf{R}+a\bm{\hat{\alpha}}} = 1\\
    l_{AB},\,\,\text{for}\,\,\sigma_{\mathbf{R}} \neq \sigma_{\mathbf{R}+a\bm{\hat{\alpha}}}\\
    l_{BB},\,\,\text{for}\,\,\sigma_{\mathbf{R}} = \sigma_{\mathbf{R}+a\bm{\hat{\alpha}}} = -1,
    \end{cases}
\end{equation}
where $a$ is the lattice constant and $\bm{\hat{\alpha}}$ is a unit
bond vector.  We take $l_{BB}<l_{AA}$ and adopt the simple mixing rule
$l_{AB}=(l_{AA}+l_{BB})/2$. The lattice mismatch $\Delta =
(l_{AA}-l_{BB})/2$ will serve as our basic unit of length.

Both the atoms' identities and their displacements ($\mathbf{u}_{\mathbf{R}}$) away from ideal
lattice positions fluctuate according to a
Boltzmann distribution $P(\{\mathbf{u}_{\mathbf{R}}\},\{\sigma_{\mathbf{R}}\})
\propto e^{-\beta \mathcal{H}}$,
where $T = (k_B\beta)^{-1}$ is
temperature and
$\mathcal{H}(\{\mathbf{u}_{\mathbf{R}}\},\{\sigma_{\mathbf{R}}\})$ is the
energy of a given configuration.  The net displacement
$\sum_{\mathbf{R}}\mathbf{u}_{\mathbf{R}}=0$ and the net fraction of $A$ atoms
$c=(2 N)^{-1}\sum_{\mathbf{R}} (\sigma_{\mathbf{R}}+1)$ are both
implicitly held fixed. Fluctuations in the lattice constant $a$ (at
zero external pressure), however, are included in the ensemble we
consider; for large systems and small lattice mismatch, this freedom
primarily allows the macroscopic geometry to adapt to the imposed
composition, $a \approx l_{AB} + \Delta (2 c-1) + O(N^{-1/2})$.  The
free energy $F(c)$ for this ensemble encodes the model's response to
changing proportions of atom types, and in particular its phase
transitions.

Deviations of bond distances away from their locally preferred lengths
incur energy that grows quadratically,
\begin{equation}
\mathcal{H}=\frac{K}{4}\sum_{\mathbf{R},\bm{\hat{\alpha}}}\left[|a\bm{\hat{\alpha}}
+\mathbf{u}_{\mathbf{R}}-\mathbf{u}_{\mathbf{R}+a\bm{\hat{\alpha}}}|-l(\sigma_{\mathbf{R}}
,\sigma_{\mathbf{R}+a\bm{\hat{\alpha}}})\right]^2,
\label{eq:hamil}
\end{equation}
where $K$ is a positive constant that sets the natural energy scale
$\epsilon = K \Delta^2/8$. All energies and lengths will henceforth be
expressed in units of $\epsilon$ and $\Delta$, respectively. The ground states of
Eq. \ref{eq:hamil} clearly occur in the absence of heterogeneity,
i.e., $c=0$ or $c=1$. At intermediate composition, fixed connectivity
prevents the collection of bonds from simultaneously attaining their
preferred lengths.  We have explored the resulting compositional
correlations analytically using small-mismatch and mean-field
approximations, and also numerically using MC simulation.

\begin{figure}
	\centering
	\includegraphics[width=\linewidth]{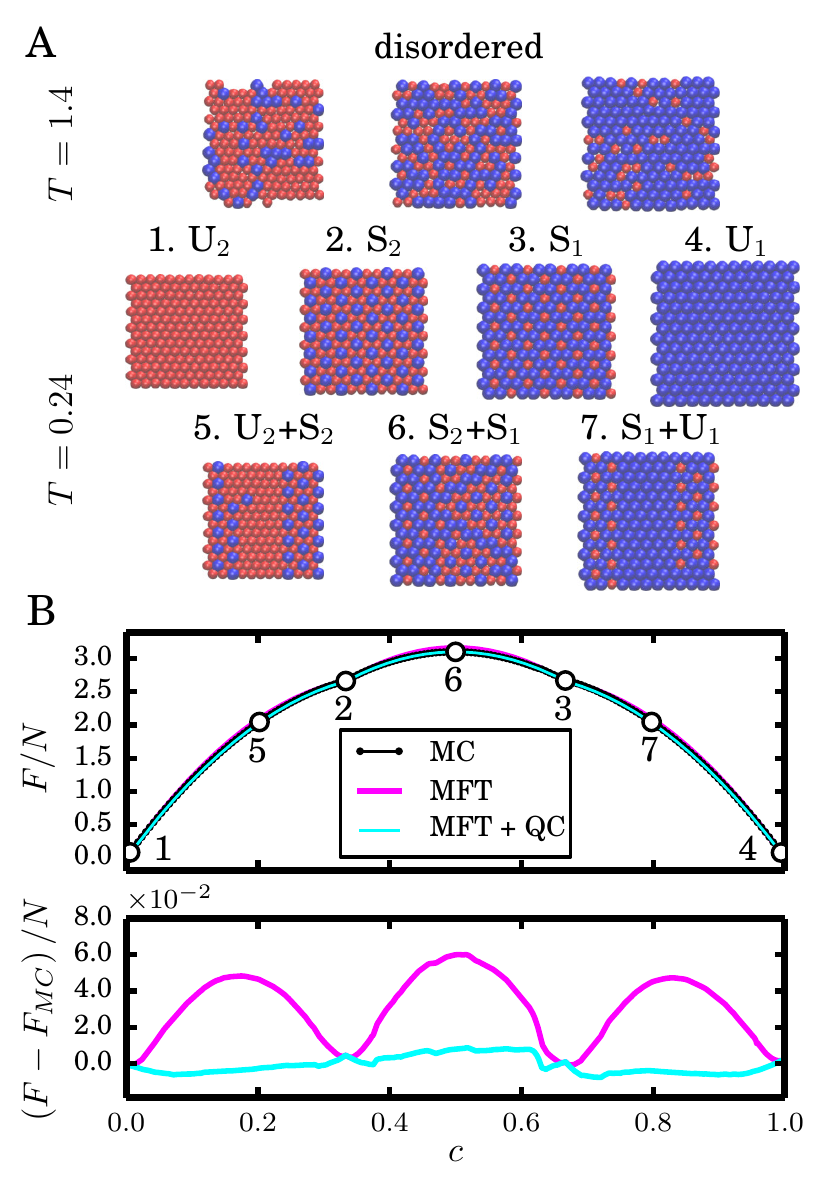}
	\caption{Monte Carlo (MC) simulation results for the elastic
		model in Eq. 2. \textbf{A}: Survey of configurations
		exemplifying the disordered, unstructured (U$_1$ and U$_2$),
		and superlattice (S$_1$ and S$_2$) phases. Blue and red
		spheres represent A and B atoms, respectively.  \textbf{B}:
		Free energy per particle $F(c)/N$ as a function of
		composition $c$ at $T=0.24$. Circles numbered 1 to 7 refer
		to the corresponding configurations in \textbf{A}. Results
		are shown in black for MC sampling, in pink for the mean
		field theory (MFT) of Eq. 5, and in blue for application of
		the quadratic construction (QC, Eq. 7) to mean-field
		thermodynamics. Lower panel shows the difference between MFT
		and MC results (pink), and the difference between MFT+QC and
		MC (blue).
	}
	\label{fig:mc}
\end{figure}

At high temperature, equilibrium states of this model are
macroscopically uniform but exhibit suggestive microscopic
correlations.  A few such disordered configurations, selected randomly
from MC simulations, are shown in the top row of
Fig. \ref{fig:mc}A. For nearly pure mixtures at modest $T$ ($\approx 1.4$,) defects
cluster in space, but not compactly. Motifs of microscopically
alternating composition are even more evident at intermediate net
composition, where typical equilibrium states resemble
interpenetrating networks of $A$ and $B$ atoms.  At low temperature
these structural tendencies produce four phases. The ``superlattice''
phases S$_1$ and S$_2$ feature periodic modulation of atom types with
wavelengths on the order of a single lattice spacing. In the vein of
previous studies of modulated order \cite{Landau1983, DeSimone1985}
we characterize these phases by their average composition on three
distinct sublattices.
In S$_1$ two sublattices are enriched in atom type $A$,
while the third is enriched in type $B$.  Roles of $A$ and $B$ are
reversed in S$_2$. The ideal forms of these phases, where the net
composition per site $2c_{\gamma}-1=(3/N)\sum_{\mathbf{R}}^{\rm (\gamma)}
\sigma_{\mathbf{R}}$ is $\pm 1$ on each sublattice $\gamma$, occur at $c=1/3$
and $c=2/3$.  In the ``unstructured'' phases U$_1$ and U$_2$, whose
zero-temperature forms are compositionally pure, the average
composition is independent of sublattice.
Previous work anticipated the appearance of modulated order phases
like S$_1$ and S$_2$ \cite{Fratzl1995, Fratzl1996}, but not their
competition with unstructured phases.

The emergence of superlattice phases as temperature decreases at
intermediate composition involves a breaking of symmetry between $A$-
and $B$-rich states. This symmetry is suggested by the form of
Eq.~\ref{eq:hamil}, but not precisely implied. Despite its
Hookean form, $\mathcal{H}$ is an anharmonic function of atomic
displacements, with nonlinearities of order $\Delta/a$ that
favor one atom type (B) for all $c\neq0,1$. The critical
point for superlattice ordering should thus occur at a value of $c$
below $1/2$. MC simulations suggest
continuous symmetry breaking very near $c=1/2$, even for the
substantial lattice mismatch $\Delta/a=0.15$, indicating that
nonlinearities in $\mathcal{H}$ are intrinsically weak in effect \cite{Note1}.

MC sampling further reveals states of coexistence among these four
phases, as depicted in the bottom row of
Fig. \ref{fig:mc}A. Specifically, S$_1$ and S$_2$ coexist at low temperature
over a range of composition centered near $c=1/2$. Coexistence between
S$_1$ and U$_1$, and between S$_2$ and U$_2$, are also observed. But under no
conditions do simulations exhibit coexistence between U$_1$ and U$_2$.

The usual quantitative signature of phase separation is a subextensive
non-convexity in the corresponding free energy,
i.e., a barrier of $O(N^{-1/2})$ in $F(c)/N$ as a function of $c$
that approaches the convex envelope
in the thermodynamic limit.  The free energies $F_{\rm MC}(c)$ we have
determined from simulation (using methods of umbrella sampling and
histogram reweighting \cite{FrenkelDSmitB2001, Torrie1977, Kumar1992}) do not follow this expectation. Specifically,
plots of $F_{\rm MC}(c)/N$ in Fig. \ref{fig:mc}B show non-convex
regions that persist as $N$ becomes large \cite{SM}.
We will argue that this behavior is
generic to the coexistence of geometrically mismatched solids with a
fixed macroscopic shape, and that the resulting negative curvature of $F(c)$ is simply related to their elastic properties.

For atom types that differ only slightly in size, $\Delta/a \ll 1$, the
energy $\mathcal{H}$ is approximately quadratic in the displacement
field $\mathbf{u}_{\mathbf{R}}$. Mechanical fluctuations in this
Gaussian limit can be integrated out exactly \cite{Fratzl1995, SM}, yielding
marginal statistics of the composition field that corresponds to a
Boltzmann distribution with effective energy
\begin{equation}
  \mathcal{H}_{\text{eff}}(\{\sigma_{\mathbf{R}}\}) =
  \frac{1}{2N}\sum_{\mathbf{q}}\tilde{V}_{\mathbf{q}}|\tilde{\sigma}_{\mathbf{q}}|^2
= \frac{1}{2}\sum_{\mathbf{R},\mathbf{R}'\neq\mathbf{R}} \sigma_{\mathbf{R}} V_{\mathbf{R}-\mathbf{R}'}
\sigma_{\mathbf{R}'},
\label{eq:eff_hamil}
\end{equation}
where
$\tilde{f}_{\mathbf{q}}=\sum_{\mathbf{R}}f_{\mathbf{R}}e^{-i\mathbf{q}\cdot\mathbf{R}}$
denotes the Fourier transform of a generic function $f_{\mathbf{R}}$.
The effective interaction potential $V_{\mathbf{R}}$ for compositional
fluctuations has Fourier components that depend smoothly on wavevector
$\mathbf{q}$ at all finite wavelengths:
\begin{equation}
\tilde{V}_{\mathbf{q}} =
\frac{4\left(2\cos{\frac{q_xa}{2}}\cos{\frac{\sqrt{3}q_ya}{2}}+\cos{q_xa}-3\right)^2}{\left(\cos{q_xa}-2\right)\left(4\cos{\frac{q_xa}{2}}\cos{\frac{\sqrt{3}q_ya}{2}}-3\right)+\cos{\sqrt{3}q_ya}},
\end{equation}
    where $x$ and $y$ indicate Cartesian components.
$\tilde{V}_{\mathbf{q}}$ vanishes abruptly at $\mathbf{q}=0$, with
important implications for open ensembles in which $c$ can vary; here,
at fixed net composition, the value of $\tilde{V}_0$ is irrelevant.

\begin{figure}
	\centering
	\includegraphics[width=\linewidth]{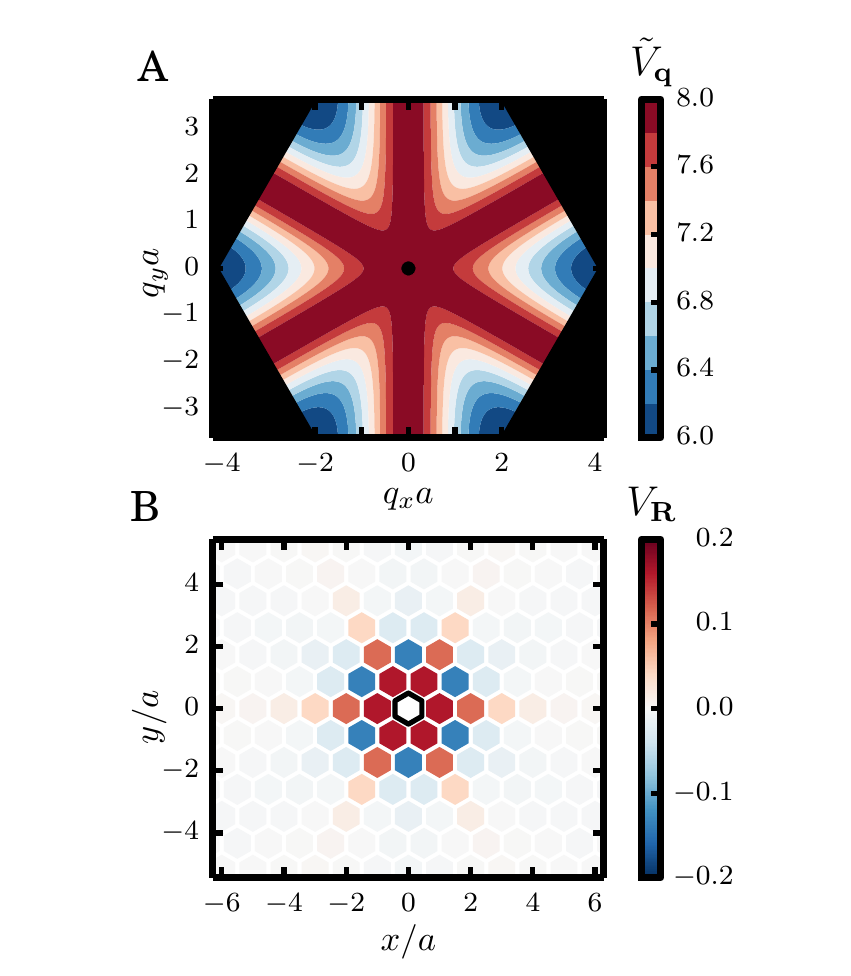}
	\caption{ Effective pair potential $V$ for the composition
          field in the small-mismatch approximation of Eq. 3.
          \textbf{A}: Reciprocal space representation
          $\tilde{V}_{\bm{q}}$, plotted in the first Brillouin zone.
          The black dot in the center indicates the discontinuity at
          $\bm{q}=\bm{0}$, where $\tilde{V}_{\bm{0}}=0$.  \textbf{B}:
          Effective interaction between an A (or B) atom at the origin
          (marked by the outlined hexagon) and another A (or B) atom
          at $\mathbf{R}$. Mixed interactions between A and B have
          opposite sign.
        }
	\label{fig:pot}
\end{figure}

Fig. \ref{fig:pot} shows the effective compositional potential in both
real- and reciprocal-space representations. Like the result of
Ref. \cite{Fratzl1995} for more complicated mechanical coupling on a
square lattice, $\tilde{V}_{\mathbf{q}}$ has local minima near the
boundary of the first Brillouin zone. Periodic variations in
composition are thus least costly at microscopic wavelengths and along
particular lattice directions, echoing the stability of superlattice
phases observed in simulations. The modulated microstructure of these
phases is suggested even more strongly by the dependence of
$V_{\mathbf{R}}$ on atom separation, which we obtain by numerical
inversion of the Fourier transform. Elastic interactions clearly
disfavor the placement of defects on neighboring lattice sites \cite{SM}.

The effective Hamiltonian
$\mathcal{H}_{\text{eff}}(\{\sigma_{\mathbf{R}}\})$ for compositional
fluctuations can serve as the basis for a simple MFT. Following standard treatments
\cite{Chandler1987, DeSimone1985}, we consider a reference system of
noninteracting spins in an external field that may differ among the
three sublattices. Variational optimization of this reference system
yields a set of self-consistent equations for the average compositions
$c_{\gamma}$ on sublattices $\gamma=1,2,3$,
\begin{equation}
2c_{\gamma}-1 = \tanh{\beta \left(\mu - 2\sum_{\delta=1}^3(2c_{\delta}-1)
	J_{\gamma\delta}\right)},
\label{eq:self_consistent}
\end{equation}
where $\mu$ is a Lagrange multiplier enforcing the constraint
$c=\sum_{\gamma}c_{\gamma}/3$, and
\begin{equation}
J_{\gamma\delta} = \frac{3}{N}\sum_{\mathbf{R}}{}^{(\gamma)}\sum_{\mathbf{R}'\neq\mathbf{R}}{}^{(\delta)}V_{\mathbf{R},\mathbf{R}'}
\end{equation}
describes the net coupling between sublattices $\gamma$ and $\delta$.

We solve Eq.~\ref{eq:self_consistent} numerically to determine an
estimate $F_{\rm MFT}(c)$ for the free energy. This mean-field
approximation successfully captures some of the general features of our
simulation results, particularly at low temperature. For the example
plotted in Fig. \ref{fig:mc}B, discrepancies are small over the entire
range of $c$, and significant only where simulations show two phases
coexisting in similar proportions. Since the states considered in MFT
are macroscopically uniform by construction, a failure to describe
phase equilibrium is expected.  From such a theory of uniform states,
assessing the thermodynamics of coexistence would typically proceed by
Maxwell construction, removing non-convex regions of $F_{\rm MFT}(c)$
that usually signal instability to the formation of interfaces. For a
case in which the true free energy is non-convex, a different procedure
is clearly needed. Here, we must specifically acknowledge an extensive
thermodynamic penalty to accommodate domains with differing lattice
constants in a rectangular macroscopic geometry. 

Linear elasticity theory associates an energy $E=Y(L-L_0)^2$ with
deforming a solid from its natural length $L_0$ to a length $L$, where
$Y$ is Young's modulus
\cite{Lifshitz1986,PhillipsRob;KondevJane;Theriot2009}.  From this
rule we can estimate the cost of phase coexistence in a
lattice-mismatched solid. Consider two phases with compositions $c_1$
and $c_2$, whose macroscopically uniform realizations have free
energies per particle $f(c_1)$ and $f(c_2)$. In the Supplemental Material \cite{SM} we estimate the
free energy of a solid in which domains
of these phases coexist at net composition $c$:
\begin{align*}
  \frac{F_{\text{coex}}(c_1,c_2;c)}{N}
  = f(c_1)-\frac{\Delta c_1}{\Delta c_2-\Delta c_1}\Delta f - Y\Delta l^2\Delta c_1 \Delta c_2.\label{eq:free_energy} \numberthis
\end{align*}
Here, $\Delta c_j = c_j-c$, $\Delta f = f(c_2)-f(c_1)$, $\Delta
l = l(c_1) - l(c_2)$, and $l(c_j)$ is the energy-minimizing unit cell
length for composition $c_j$.

Absent lattice mismatch ($\Delta l = 0$), minimizing
Eq.~\ref{eq:free_energy} with respect to $c_1$ and $c_2$ (at fixed
$c$) corresponds to the conventional double-tangent construction.  For
$\Delta l \neq 0$, coexistence instead entails a free energy that
connects points $(c_1^*,f(c_1^*))$ and $(c_2^*,f(c_2^*))$
in the $c$-$f$ plane with a parabola of curvature $\kappa_{\rm
  coex}=-Y\Delta l^2$.
We term this procedure the ``quadratic construction'' (QC) \cite{SM}.

Applying the QC to our MFT estimate $F_{\text{MFT}}(c)$,
correspondence with MC results can be greatly improved. In the case of Fig. 1B,
mean-field predictions for $F(c)$ deviate from simulations by less
than 1\%, comparable to random sampling error. This excellent
agreement emphasizes a predominance of macroscopically heterogeneous
states in the temperature range $T \lesssim 0.4$, despite
the non-convexity of $F(c)$.  We attribute this agreement to the
appreciable spatial range of $V_{\mathbf{R}}$, which includes
substantial coupling between sites separated by several lattice
spacings. The low-$\bq$ form of $\tilde{V}_{\mathbf{q}}$, which varies
quadratically with $q$ to lowest order, suggests an
eventual failure of MFT near criticality
\cite{Fisher1972,SM}. Quantitative agreement indeed deteriorates with
increasing temperature, and above $T \approx 0.46$ the fluctuations
neglected by MFT influence phase behavior even qualitatively.  The
phase diagram for our elastic model, as determined from MC simulations
\cite{SM} and plotted in Fig. 3B, is equivalent in form to a spin model on the same lattice with
couplings that resemble $V_{\mathbf{R}}$ at short range
\cite{Landau1983}.  In contrast to predictions of MFT (see Fig. 3A,) (i) the loss of
superlattice order upon heating is continuous, with critical
properties belonging to the three-state Potts model universality
class, and (ii) in the temperature range $T\approx 0.46$ to $T\approx
0.56$, phases S$_1$ and S$_2 $ are separated by a line of Kosterlitz-Thouless
critical points. Away from these exotic features, first order
transitions are well described by Eqs. \ref{eq:self_consistent} and \ref{eq:free_energy}. The absence of a first
order transition between unstructured phases U$_1$ and U$_2$ is also
captured by MFT and the QC, which manifest an energetic instability
for this scenario \cite{SM}.

\begin{figure}
	\centering
	\includegraphics[width=\linewidth]{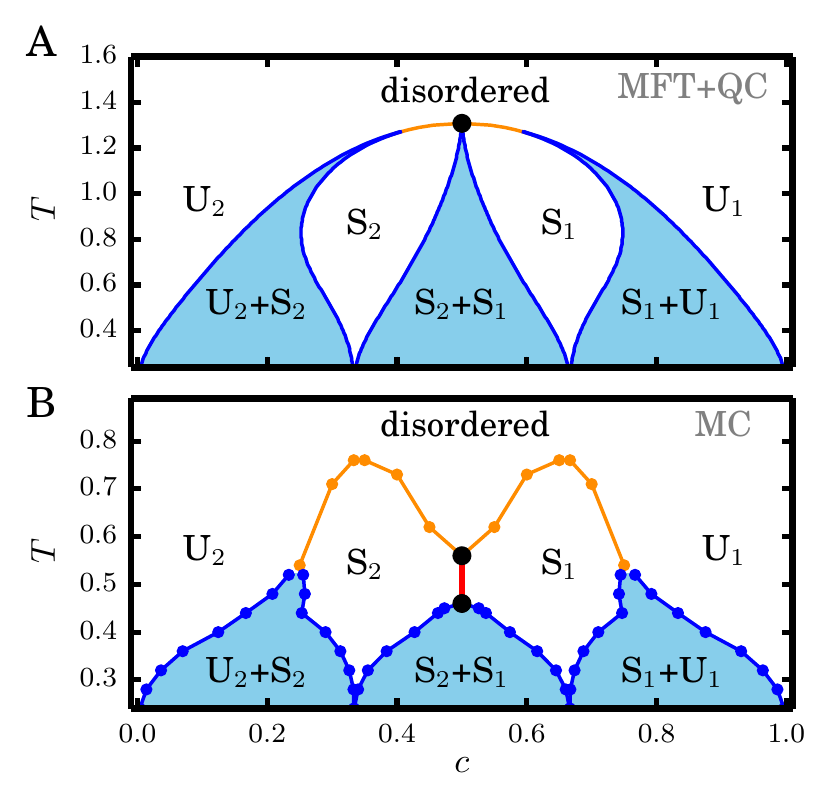}
	\caption{Phase diagram for our elastic model in the plane of
          temperature and composition.  \textbf{A}: Mean-field
          prediction resulting from the quadratic construction of
          Eq. \ref{eq:free_energy}. Black circle indicates a critical point at $T\approx
          1.3$; elsewhere, lines indicate first-order transitions.
          Orange lines separate the disordered phase from superlattice
          phases S$_1$ or S$_2$. Blue lines bound coexistence regions,
          which are shaded in light blue.  \textbf{B}: Numerically
          exact results from Monte Carlo sampling. In this case, the
          disordered-to-superlattice transitions (orange lines) are
          continuous. A line of Kosterlitz-Thouless critical points
          between $T_c^{\text{lower}}\approx0.46$ and
          $T_c^{\text{upper}}\approx0.56$ is shown in red.}
	\label{fig:phase_diagram}
\end{figure}

Our results demonstrate that lattice mismatch can generate more
nuanced thermodynamic behaviors than was previously appreciated. They
also indicate a central importance of lattice geometry and boundary
conditions. The modulated order of phases S$_1$ and S$_2$ owes its stability
to the fixed macroscopic shape implied by periodic boundary
conditions. Such a constraint on boundary shape could arise in real
systems from strong interactions that bind a nanocrystal to a
substrate, a notion consistent with the observation of
stable Cu superlattices within two-dimensional Bi$_2$Se$_3$
nanocrystals \cite{Buha2017}. It could also be imposed by core-shell interactions in
hetero-nanostructures. Core/shell arrangements, moreover, are
natural intermediates in the course of exchange reactions that
proceed most rapidly at surface sites \cite{Groeneveld2013}.

The precise form of the phase diagram in Fig. 3B is likely
specific to the dimensionality and lattice symmetry of the elastic
model we have studied. Several of its interesting features, however,
we expect to be general for heterostructured solids under appropriate
boundary conditions. A tendency for modulated order, for example, is
evident in three-dimensional systems explored previously
\cite{Gupta2001} and in exploratory simulations described in SM \cite{SM}.
Thermodynamic potentials with indefinite convexity, and their
implications for phase coexistence, are similarly anticipated as
generic consequences of the elastic forces attending lattice mismatch.
Testing these predictions in the laboratory may be most
straightforward for materials that can be manipulated more readily
than the internal structure of nanocrystals, for instance assemblies
of DNA-coated nanoparticles \cite{Gabrys2018} or spin-crossover
compounds \cite{Enachescu2015,Nakada2012,Nicolazzi2012,Nishino2007a,Slimani2013,Spiering1989}, where elasticity is known
to play a significant role.

We thank Jaffar Hasnain for stimulating conversations. This work was supported by National Science Foundation (NSF) grant CHE-1416161. This research also used resources of the National Energy Research Scientific Computing Center (NERSC), a U.S. Department of Energy Office of Science User Facility operated under Contract No. DE-AC02-05CH11231. 

\bibliography{final_refs.bib}

\ifarXiv


\section*{Supplementary material}

# Supplemental Material for "Consequences of lattice mismatch for phase equilibrium in heterostructured solids"

Layne B. Frechette and Phillip L. Geissler

*Department of Chemistry, University of California, Berkeley, California 94720, USA and*
*Erwin Schrödinger Institute for Mathematics and Physics,*
*University of Vienna, Boltzmanngasse 9, Wien 1090, Austria*

Christoph Dellago

*Faculty of Physics, University of Vienna, Boltzmanngasse 5, Wien 1090, Austria and*
*Erwin Schrödinger Institute for Mathematics and Physics,*
*University of Vienna, Boltzmanngasse 9, Wien 1090, Austria*

(Dated: June 18, 2019)

## SMALL MISMATCH APPROXIMATION

In order to make our Hamiltonian analytically tractable, we make the approximation that the mismatch is small. To begin, note that the natural bond length, Eq. 1 in the main text, can be rewritten as:

$$l(\sigma_{\mathbf{R}}, \sigma_{\mathbf{R}'}) = l_{AB} + \frac{\Delta}{2}(\sigma_{\mathbf{R}} + \sigma_{\mathbf{R}'}). \tag{1}$$

We then rewrite the full Hamiltonian, Eq. 2 in the main text, as:

$$\mathcal{H}/\epsilon = \bar{\mathcal{H}} = 2\sum_{\mathbf{R},\hat{\boldsymbol{\alpha}}} \left[ \frac{a}{\Delta} \left| \hat{\boldsymbol{\alpha}} + \left(\frac{\Delta}{a}\right)(\bar{\mathbf{u}}_{\mathbf{R}} - \bar{\mathbf{u}}_{\mathbf{R}+a\hat{\boldsymbol{\alpha}}}) \right| - \left( \frac{l_{AB}}{\Delta} + \frac{1}{2}(\sigma_{\mathbf{R}} + \sigma_{\mathbf{R}+a\hat{\boldsymbol{\alpha}}}) \right) \right]^2. \tag{2}$$

Here we have defined a scaled Hamiltonian $\bar{\mathcal{H}} = \mathcal{H}/\epsilon$ and scaled displacements $\bar{\mathbf{u}}_{\mathbf{R}} = \mathbf{u}_{\mathbf{R}}/\Delta$. Now, let $\boldsymbol{\delta} = (\Delta/a)(\bar{\mathbf{u}}_{\mathbf{R}} - \bar{\mathbf{u}}_{\mathbf{R}+a\hat{\boldsymbol{\alpha}}})$. Then,

$$|\hat{\boldsymbol{\alpha}} + \boldsymbol{\delta}|^2 = |\hat{\boldsymbol{\alpha}}|^2 + 2\hat{\boldsymbol{\alpha}} \cdot \boldsymbol{\delta} + |\boldsymbol{\delta}|^2$$
$$= 1 + 2\hat{\boldsymbol{\alpha}} \cdot \boldsymbol{\delta} + \mathcal{O}(\delta^2)$$
$$\implies |\hat{\boldsymbol{\alpha}} + \boldsymbol{\delta}| = \sqrt{1 + 2\hat{\boldsymbol{\alpha}} \cdot \boldsymbol{\delta} + \mathcal{O}(\delta^2)}. \tag{3}$$

For small $x$, we have $\sqrt{1+x} = 1 + \frac{x}{2} + \mathcal{O}(x^2)$, so:

$$|\hat{\boldsymbol{\alpha}} + \boldsymbol{\delta}| \approx 1 + \hat{\boldsymbol{\alpha}} \cdot \boldsymbol{\delta}. \tag{4}$$

Therefore,

$$\bar{\mathcal{H}} \approx 2\sum_{\mathbf{R},\hat{\boldsymbol{\alpha}}} \left[ \frac{a}{\Delta} + \frac{a}{\Delta}\hat{\boldsymbol{\alpha}} \cdot \boldsymbol{\delta} - \frac{l_{AB}}{\Delta} - \frac{1}{2}(\sigma_{\mathbf{R}} + \sigma_{\mathbf{R}+a\hat{\boldsymbol{\alpha}}}) \right]^2$$
$$= 2\sum_{\mathbf{R},\hat{\boldsymbol{\alpha}}} \left[ \hat{\boldsymbol{\alpha}} \cdot (\bar{\mathbf{u}}_{\mathbf{R}} - \bar{\mathbf{u}}_{\mathbf{R}+a\hat{\boldsymbol{\alpha}}}) + \frac{a - l_{AB}}{\Delta} - \frac{1}{2}(\sigma_{\mathbf{R}} + \sigma_{\mathbf{R}+a\hat{\boldsymbol{\alpha}}}) \right]^2. \tag{5}$$

We define the quantity:

$$\delta\sigma_{\mathbf{R}} = \sigma_{\mathbf{R}} - \frac{a - l_{AB}}{\Delta}, \tag{6}$$

allowing us to write our small-mismatch Hamiltonian as:

$$\bar{\mathcal{H}} \approx 2\sum_{\mathbf{R},\hat{\boldsymbol{\alpha}}} \left( \hat{\boldsymbol{\alpha}} \cdot (\bar{\mathbf{u}}_{\mathbf{R}} - \bar{\mathbf{u}}_{\mathbf{R}+a\hat{\boldsymbol{\alpha}}}) - \frac{1}{2}(\delta\sigma_{\mathbf{R}} + \delta\sigma_{\mathbf{R}+a\hat{\boldsymbol{\alpha}}}) \right)^2. \tag{7}$$

From the above expression it is clear that, in the small-mismatch limit, our Hamiltonian is quadratic in both displacement and spin variables, and that the displacements and spins are linearly coupled. We drop the overbar on the displacements and Hamiltonian in what follows.

## FOURIER SPACE AND THE q = 0 MODE

Translational invariance implies that we can diagonalize our Hamiltonian by writing it in Fourier space. We first write the displacement and spin fields in terms of their Fourier coefficients:

$$\mathbf{u_R} = \frac{1}{N} \sum_{\mathbf{q}} \tilde{\mathbf{u}}_\mathbf{q} e^{-i\mathbf{q}\cdot\mathbf{R}} \tag{8a}$$

$$\delta\sigma_\mathbf{R} = \frac{1}{N} \sum_{\mathbf{q}} \delta\tilde{\sigma}_\mathbf{q} e^{-i\mathbf{q}\cdot\mathbf{R}}, \tag{8b}$$

where the sum is taken over values within the first Brillouin zone, specifically $\mathbf{q} = n_1 \mathbf{b}_1/N_1 + n_2 \mathbf{b}_2/N_2$, where $N_1 N_2 = N$; $n_1 = 0, 1, \ldots, N_1 - 1$; $n_2 = 0, 1, \ldots, N_2 - 1$; and the reciprocal lattice vectors for the triangular lattice are given by [1]:

$$\mathbf{b}_1 = \frac{2\pi}{\sqrt{3}a}\left(\sqrt{3}, -1\right) \tag{9a}$$

$$\mathbf{b}_2 = \frac{4\pi}{\sqrt{3}a}(0, 1). \tag{9b}$$

Inserting equations 8a and 8b into equation 7, and using the relation:

$$\sum_{\mathbf{R}} e^{-i(\mathbf{q}+\mathbf{q}')\cdot\mathbf{R}} = N\delta_{-\mathbf{q},\mathbf{q}'}, \tag{10}$$

where $\delta_{-\mathbf{q},\mathbf{q}'}$ denotes the Kronecker delta function, we obtain a Fourier-space version of our Hamiltonian:

$$\mathcal{H} = \frac{2}{N} \sum_{\mathbf{q}} \left[\tilde{\mathbf{u}}_\mathbf{q} \cdot \mathbf{F}_\mathbf{q} \cdot \tilde{\mathbf{u}}_{-\mathbf{q}} + \frac{1}{4} g_\mathbf{q} \delta\tilde{\sigma}_\mathbf{q} \delta\tilde{\sigma}_{-\mathbf{q}} - \frac{1}{2} \mathbf{h}_\mathbf{q} \cdot (\tilde{\mathbf{u}}_\mathbf{q} \delta\tilde{\sigma}_{-\mathbf{q}} - \tilde{\mathbf{u}}_{-\mathbf{q}} \delta\tilde{\sigma}_\mathbf{q})\right]. \tag{11}$$

The quantities $\mathbf{F}$, $\mathbf{h}$, and $g$ encode information about the lattice structure:

$$\mathbf{F}_\mathbf{q} = \sum_{\hat{\boldsymbol{\alpha}}} f_{\hat{\boldsymbol{\alpha}},\mathbf{q}} \hat{\boldsymbol{\alpha}}\hat{\boldsymbol{\alpha}} \tag{12a}$$

$$\mathbf{h}_\mathbf{q} = \sum_{\hat{\boldsymbol{\alpha}}} h_{\hat{\boldsymbol{\alpha}},\mathbf{q}} \hat{\boldsymbol{\alpha}} \tag{12b}$$

$$g_\mathbf{q} = \sum_{\hat{\boldsymbol{\alpha}}} g_{\hat{\boldsymbol{\alpha}},\mathbf{q}}, \tag{12c}$$

where:

$$f_{\hat{\boldsymbol{\alpha}},\mathbf{q}} = \left(e^{-ia\mathbf{q}\cdot\hat{\boldsymbol{\alpha}}} - 1\right)\left(e^{ia\mathbf{q}\cdot\hat{\boldsymbol{\alpha}}} - 1\right) \tag{13a}$$

$$h_{\hat{\boldsymbol{\alpha}},\mathbf{q}} = \left(e^{-ia\mathbf{q}\cdot\hat{\boldsymbol{\alpha}}} - 1\right)\left(e^{ia\mathbf{q}\cdot\hat{\boldsymbol{\alpha}}} + 1\right) \tag{13b}$$

$$g_{\hat{\boldsymbol{\alpha}},\mathbf{q}} = \left(e^{-ia\mathbf{q}\cdot\hat{\boldsymbol{\alpha}}} + 1\right)\left(e^{ia\mathbf{q}\cdot\hat{\boldsymbol{\alpha}}} + 1\right). \tag{13c}$$

It is revealing to single out the longest-wavelength ($\mathbf{q} = 0$) component of the Hamiltonian. In particular, we write:

$$\mathcal{H} = \mathcal{H}_{\mathbf{q}\neq 0} + \mathcal{H}_{\mathbf{q}=0}, \tag{14}$$

where $\mathcal{H}_{\mathbf{q}\neq 0}$ is simply Eq. 11 with the $\mathbf{q} = 0$ term excluded, and $\mathcal{H}_{\mathbf{q}=0}$ is:

$$\mathcal{H}_{\mathbf{q}=0} = \frac{2Z}{N} \delta\tilde{\sigma}_0^2, \tag{15}$$

since $f_{\hat{\boldsymbol{\alpha}},\mathbf{q}}$ and $h_{\hat{\boldsymbol{\alpha}},\mathbf{q}}$ are both zero when $\mathbf{q} = 0$, and $g_0 = 2Z$, where $Z$ is the coordination number of the lattice. Now, by taking the Fourier transform of $\delta\sigma_\mathbf{R} = \sigma_\mathbf{R} - (a - l_{AB})/\Delta$, we obtain:

$$\delta\tilde{\sigma}_\mathbf{q} = \tilde{\sigma}_\mathbf{q} - N\delta_{\mathbf{q},0}\left(\frac{a - l_{AB}}{\Delta}\right), \tag{16}$$

where $\tilde{\sigma}_{\mathbf{q}} = \sum_{\mathbf{R}} \sigma_{\mathbf{R}} e^{i\mathbf{q}\cdot\mathbf{R}}$. (This fact allows us to drop the $\delta$ in front of $\sigma$ for all but the $\mathbf{q} = 0$ mode.) Using the fact that $\tilde{\sigma}_0 = \sum_{\mathbf{R}} \sigma_{\mathbf{R}} = N\langle\sigma\rangle$, where $\langle\cdots\rangle$ denotes a spatial average, we can substitute Eq. 16 into Eq. 15 to obtain:

$$\mathcal{H}_{\mathbf{q}=0} = 2ZN\left(\langle\sigma\rangle - \left(\frac{a - l_{AB}}{\Delta}\right)\right)^2. \tag{17}$$

Thus, if the composition $c$ is fixed (so $\langle\sigma\rangle = 2c - 1$), the lattice constant $a$ will exhibit harmonic fluctuations about the minimum-energy value:

$$a = l_{AB} + \Delta\langle\sigma\rangle. \tag{18}$$

However, the energy penalty for these fluctuations is manifestly extensive and so such fluctuations can be neglected in the thermodynamic limit. We thus see that the net composition effectively sets the system's volume. Finite-sized simulations exhibit fluctuations in $a$ that are consistent with this prediction.

## EFFECTIVE HAMILTONIAN

We can integrate out the displacement field to obtain an effective Hamiltonian depending only on the spin variables. First let us write the fields in terms of their real and imaginary parts:

$$\tilde{\mathbf{u}}_{\mathbf{q}} = \mathbf{a}_{\mathbf{q}} + i\mathbf{b}_{\mathbf{q}} \tag{19a}$$
$$\tilde{\mathbf{u}}_{-\mathbf{q}} = \mathbf{a}_{\mathbf{q}} - i\mathbf{b}_{\mathbf{q}} \tag{19b}$$
$$\tilde{\sigma}_{\mathbf{q}} = c_{\mathbf{q}} + id_{\mathbf{q}} \tag{19c}$$
$$\tilde{\sigma}_{-\mathbf{q}} = c_{\mathbf{q}} - id_{\mathbf{q}}. \tag{19d}$$

Plugging these into the Hamiltonian yields:

$$\mathcal{H}_{\mathbf{q}\neq 0} = \frac{2}{N}\sum_{\mathbf{q}\neq 0}\left[\mathbf{a}_{\mathbf{q}}\cdot\mathbf{F}_{\mathbf{q}}\cdot\mathbf{a}_{\mathbf{q}} + \mathbf{b}_{\mathbf{q}}\cdot\mathbf{F}_{\mathbf{q}}\cdot\mathbf{b}_{\mathbf{q}} + \frac{1}{4}(c_{\mathbf{q}}^2 + d_{\mathbf{q}}^2)g_{\mathbf{q}} - i\Delta\mathbf{h}_{\mathbf{q}}\cdot(c_{\mathbf{q}}\mathbf{b}_{\mathbf{q}} - d_{\mathbf{q}}\mathbf{a}_{\mathbf{q}})\right]. \tag{20}$$

Now we need to integrate out the fields $\mathbf{a}_{\mathbf{q}}$ and $\mathbf{b}_{\mathbf{q}}$. This is equivalent (up to an unimportant constant) to minimizing the Hamiltonian with respect to $\mathbf{a}_{\mathbf{q}}$ and $\mathbf{b}_{\mathbf{q}}$, since their fluctuations are harmonic. Setting the derivatives of Eq. 20 with respect to $\mathbf{a}_{\mathbf{q}}$ and $\mathbf{b}_{\mathbf{q}}$ equal to zero, we find:

$$\mathbf{a}_{\mathbf{q}} = -i\frac{1}{2}d_{\mathbf{q}}\mathbf{F}_{\mathbf{q}}^{-1}\cdot\mathbf{h}_{\mathbf{q}} \tag{21}$$
$$\mathbf{b}_{\mathbf{q}} = i\frac{1}{2}c_{\mathbf{q}}\mathbf{F}_{\mathbf{q}}^{-1}\cdot\mathbf{h}_{\mathbf{q}}. \tag{22}$$

Plugging these values into our Hamiltonian and simplifying gives us an effective Hamiltonian:

$$\mathcal{H}_{\text{eff}} = \frac{1}{2N}\sum_{\mathbf{q}\neq 0}|\tilde{\sigma}_{\mathbf{q}}|^2\left(\mathbf{h}_{\mathbf{q}}\cdot\mathbf{F}_{\mathbf{q}}^{-1}\cdot\mathbf{h}_{\mathbf{q}} + g_{\mathbf{q}}\right). \tag{23}$$

where we have used the fact that $c^2 + d^2 = |\tilde{\sigma}_{\mathbf{q}}|^2$. We identify the term in parentheses as a Fourier-space effective potential:

$$\tilde{U}_{\mathbf{q}} = \mathbf{h}_{\mathbf{q}}\cdot\mathbf{F}_{\mathbf{q}}^{-1}\cdot\mathbf{h}_{\mathbf{q}} + g_{\mathbf{q}}, \tag{24}$$

so we can write the effective Hamiltonian as:

$$\mathcal{H}_{\text{eff}} = \frac{1}{2N}\sum_{\mathbf{q}\neq 0}\tilde{U}_{\mathbf{q}}|\tilde{\sigma}_{\mathbf{q}}|^2. \tag{25}$$

Instead of excluding the $\mathbf{q} = 0$ term from the sum, it is convenient to define a new effective potential:

$$\tilde{V}_{\mathbf{q}} = \begin{cases} \tilde{U}_{\mathbf{q}}, & \mathbf{q} \neq 0 \\ 0, & \mathbf{q} = 0, \end{cases} \tag{26}$$

so that our effective Hamiltonian is:

$$\mathcal{H}_{\text{eff}} = \frac{1}{2N} \sum_{\mathbf{q}} \tilde{V}_{\mathbf{q}} |\tilde{\sigma}_{\mathbf{q}}|^2, \tag{27}$$

In order to examine the features of the effective potential in more detail, we must specify a lattice structure. For the triangular lattice, the bond vectors are:

$$\hat{\boldsymbol{\alpha}}_1 = \begin{pmatrix} 1/2 \\ \sqrt{3}/2 \end{pmatrix} \tag{28a}$$

$$\hat{\boldsymbol{\alpha}}_2 = \begin{pmatrix} 1 \\ 0 \end{pmatrix} \tag{28b}$$

$$\hat{\boldsymbol{\alpha}}_3 = \begin{pmatrix} 1/2 \\ -\sqrt{3}/2 \end{pmatrix}, \tag{28c}$$

as well as their additive inverses $-\hat{\boldsymbol{\alpha}}_1$, $-\hat{\boldsymbol{\alpha}}_2$, $-\hat{\boldsymbol{\alpha}}_3$. The sum over bond vectors $\hat{\boldsymbol{\alpha}}$ includes all six of these vectors. The resulting effective potential is given by Eq. 4 in the main text. It is worth noting that the potential is discontinuous at $\mathbf{q} = 0$. We have $\lim_{\mathbf{q} \to 0} \tilde{V}_{\mathbf{q}} = 8$, but $\tilde{V}_{\mathbf{q}=0} = 0$. This results in an infinitely long-ranged contribution to the potential whose magnitude goes as $1/N$. We will explore the implications of this contribution in more detail in a subsequent paper.

## SIMULATION DETAILS

We used Monte Carlo (MC) simulations to explore the equilibrium properties of the full Hamiltonian (Eq. 2 in the main text) on the triangular lattice. Periodic boundary conditions were imposed in order to simulate bulk behavior. The numbers of unit cells in the $x$ and $y$ directions, $N_x$ and $N_y$, were chosen so as to make the dimensions $L_x$, $L_y$ of the box nearly equal. Many of simulations were carried out with $N_x = 12$, $N_y = 14$ (making the total number of atoms $N = 168$,) so that $L_x/L_y \approx 0.990$. A number of simulations were also performed at different system sizes, the largest being $N_x = 36$, $N_y = 40$, in order to assess finite-size effects. These will be described in more detail later.

Most simulations were conducted in the isothermal-isobaric (NPT) ensemble. This ensemble was sampled using two basic MC moves: spin flips and displacement moves. In both cases, an atom at lattice site $\mathbf{R}$ was selected at random. An attempt was then made to change either its spin, $\sigma_{\mathbf{R}} \to -\sigma_{\mathbf{R}}$, or its position, $\mathbf{R} + \mathbf{u}_{\mathbf{R}} = \mathbf{r}_{\mathbf{R}} \to \mathbf{r}_{\mathbf{R}} + \boldsymbol{\delta}$, where $\boldsymbol{\delta} = (\delta_x, \delta_y)$ is a two dimensional vector. The components of the vector were selected uniformly at random from an interval $[-\delta_{\max}, \delta_{\max}]$. The requisite random numbers were generated using the Mersenne Twister algorithm [2] as implemented in the GNU Scientific Library (GSL) [3]. For most of our simulations, $\delta_{\max}$ was chosen to be $0.1 l_{AA}$, although at the lowest temperatures we considered it was more efficient to use a smaller value of $0.05 l_{AA}$. After a move (either displacement or spin) was proposed, it was either accepted or rejected according to a Metropolis criterion [4]:

$$P(C \to C') = \min\left[1, e^{-\beta(\mathcal{H}(C) - \mathcal{H}(C'))}\right], \tag{29}$$

where $C$ and $C'$ represent configurations $\{\sigma_{\mathbf{R}}\}$, $\{\mathbf{r}_{\mathbf{R}}\}$ before and after the proposed move, respectively, and $\beta = 1/k_B T$. The above form of the acceptance probability ensures that the MC algorithm satisfies detailed balance and hence properly samples the equilibrium distribution at temperature $T$.

Simulation runs consisted of performing a large number of MC sweeps. A single MC sweep consisted of $N$ attempted spin flips and $N$ attempted displacement moves. Constant pressure was maintained using a standard algorithm in which attempts to change the system's volume were proposed and then accepted or rejected according to a Metropolis criterion [4]. In our simulations, proposed volume moves consisted of changing the total volume by an amount $\delta V$, selected uniformly at random from the interval $[-\delta V_{\max}, \delta V_{\max}]$. We chose $\delta V_{\max} = 0.01 V_0$, where $V_0$ is the volume at the beginning of a simulation run. Such volume moves were performed once every MC sweep.

Before collecting data from any molecular simulation, it is necessary to equilibrate the system in order to remove artifacts due to its initial preparation. In our simulations, initial configurations consisted of atoms arranged on a perfect triangular lattice, with a random distribution of spins, and with a volume consistent with the net composition. Before obtaining statistics, we equilibrated the system by running for $10^7$ MC sweeps without collecting any data. Statistics were then collected for different observables once every sweep, over $10^7$ sweeps.

A central quantity of interest is the free energy as a function of composition, $F(c)$. We used umbrella sampling [4, 5] to obtain this quantity. For calculations at temperatures $T \geq 2.0$ we biased the system with a set of hard wall window potentials spaced equally over the range of possible compositions [4]. For the system size $N_x = 12$, $N_y = 14$, we used 100 overlapping windows. Within each window, statistics were collected as described in the previous paragraph. We then used the Weighted Histogram Analysis Method (WHAM) [6] to construct the full probability distribution (and hence free energy) as a function of composition. For lower temperatures ($k_B T/\epsilon < 2.0$), large free energy barriers necessitated a somewhat more involved procedure. Here we used the inverted free energy profile obtained at $k_B T = 2.0$, in combination with a set of evenly spaced harmonic windows with spring constant $K' = 20\epsilon/\Delta^2$, as a bias potential. For $N_x = 12$, $N_y = 14$, 100 harmonic windows were used. Larger systems required more windows - for $N_x = 24$, $N_y = 28$, we used 400 windows. We again used WHAM to construct free energy profiles from this data.

The majority of our simulations were carried out with a mismatch of $\Delta = 0.1$. The MC results presented here and in the main text were obtained using this value. However, we also ran the same simulations at a much smaller value of $\Delta = 0.005$ and obtained essentially identical results.

In some cases we computed thermodynamic quantities at an explicitly fixed composition. This necessitated the use of Kawasaki (rather than Metropolis) dynamics [7]. In Kawasaki dynamics, one performs a spin *exchange* rather than a spin flip. Explicitly, two atoms are selected at random and, if their spins are different, an exchange of their values is attempted. This proposed move is accepted with the detailed balance-preserving probability of Eq. 29.

## MEAN FIELD THEORY

Our mean field theory considers our effective Hamiltonian subject to the requirement that the net composition $c$ (which can take on real values between 0 and 1) be fixed:

$$\bar{m} = 2c - 1 = \frac{1}{N} \sum_{\mathbf{R}} \sigma_{\mathbf{R}}. \tag{30}$$

In analogy with magnetic systems, we refer to $\bar{m}$ as the net magnetization. Anticipating modulated order, we write a reference Hamiltonian in which each of $n$ sublattices is subject to a different effective field [8]:

$$\mathcal{H}_0 = -\sum_{\gamma=1}^{n} h_\gamma \sum_{\mathbf{R}}^{(\gamma)} \sigma_{\mathbf{R}}, \tag{31}$$

where $h_\gamma$ is an effective field that acts only on sublattice $\gamma$, and the $\gamma$ superscript in the sum over lattice sites emphasizes that the sum is taken only over sites on sublattice $\gamma$. The corresponding reference partition function is:

$$Q_0 = \sum_{\{\sigma_{\mathbf{R}}\}}{}' e^{-\beta \mathcal{H}_0(\{\sigma_{\mathbf{R}}\})}, \tag{32}$$

where the prime symbol reminds us that the sum must respect the constraint $\frac{1}{N} \sum_{\mathbf{R}} \sigma_{\mathbf{R}} = \bar{m}$. Performing the constrained sum directly is difficult. Instead, we introduce a Lagrange multiplier $\mu$ and write:

$$Q_0 = \sum_{\{\sigma_{\mathbf{R}}\}} e^{\beta \mu \left( \sum_{\mathbf{R}} \sigma_{\mathbf{R}} - N\bar{m} \right) - \beta \mathcal{H}_0}$$

$$= e^{-\beta \mu N \bar{m}} \prod_{\gamma} \prod_{\mathbf{R}}^{(\gamma)} 2 \cosh \beta(\mu + h_\gamma)$$

$$= e^{-\beta \mu N \bar{m}} \prod_{\gamma} [2 \cosh \beta(\mu + h_\gamma)]^{N_\gamma},$$

where $N_\gamma$ is the number of sites on sublattice $\gamma$. The sublattice magnetization, defined as:

$$m_\gamma = \langle \sigma_{\mathbf{R}} \rangle^{(\gamma)} = \frac{1}{N_\gamma} \sum_{\mathbf{R}}^{(\gamma)} \sigma_{\mathbf{R}}, \tag{33}$$

is determined by the soon-to-be self-consistent equation:

$$m_\gamma = \frac{1}{N_\gamma} \frac{\partial \log Q_0}{\partial \beta h_\gamma} = \tanh \beta(\mu + h_\gamma). \tag{34}$$

Now we apply the Gibbs-Bogoliubov-Feynman estimate [9]:

$$Q_{\text{est}} = Q_0 e^{-\beta \langle \Delta \mathcal{H} \rangle_0}, \quad (35)$$

where $\Delta \mathcal{H} = \mathcal{H}_{\text{eff}} - \mathcal{H}_0$, and where $\langle \ldots \rangle_0$ indicates an average with respect to the reference Hamiltonian. To determine the optimal effective fields we maximize the logarithm of $Q_{\text{est}}$:

$$0 = \frac{\partial \log Q_{\text{est}}}{\partial \beta h_\gamma}$$
$$= \underbrace{\frac{\partial \log Q_0}{\partial \beta h_\gamma}}_{(i)} - \underbrace{\beta \frac{\partial \langle \mathcal{H}_{\text{eff}} \rangle_0}{\partial \beta h_\gamma}}_{(ii)} + \underbrace{\beta \frac{\partial \langle \mathcal{H}_0 \rangle_0}{\partial \beta h_\gamma}}_{(iii)}. \quad (36)$$

We evaluate Eq. 36 term-by-term:

$$(i): \quad \frac{\partial \log Q_0}{\partial \beta h_\gamma} = \left\langle \sum_{\mathbf{R}}{}^{(\gamma)} \sigma_{\mathbf{R}} \right\rangle_0$$
$$= \boxed{N_\gamma m_\gamma.} \quad (37)$$

Next,

$$(ii): \quad \langle \mathcal{H}_{\text{eff}} \rangle_0 = \left\langle \frac{1}{2} \sum_{\mathbf{R},\mathbf{R}' \neq \mathbf{R}} V_{\mathbf{R},\mathbf{R}'} \sigma_\mathbf{R} \sigma_{\mathbf{R}'} \right\rangle_0$$
$$= \frac{1}{2} \left\langle \sum_{\gamma,\delta} \sum_{\mathbf{R}}{}^{(\gamma)} \sum_{\mathbf{R}' \neq \mathbf{R}}{}^{(\delta)} V_{\mathbf{R},\mathbf{R}'} \sigma_\mathbf{R}^{(\gamma)} \sigma_{\mathbf{R}'}^{(\delta)} \right\rangle_0$$
$$= \frac{1}{2} \sum_{\gamma,\delta} \sum_{\mathbf{R}}{}^{(\gamma)} \sum_{\mathbf{R}' \neq \mathbf{R}}{}^{(\delta)} V_{\mathbf{R},\mathbf{R}'} \left\langle \sigma_\mathbf{R}^{(\gamma)} \right\rangle_0 \left\langle \sigma_{\mathbf{R}'}^{(\delta)} \right\rangle_0$$
$$= \sum_{\gamma,\delta} J_{\gamma\delta} m_\gamma m_\delta, \quad (38)$$

where we have defined:

$$J_{\gamma\delta} = \frac{1}{2} \sum_{\mathbf{R}}{}^{(\gamma)} \sum_{\mathbf{R}' \neq \mathbf{R}}{}^{(\delta)} V_{\mathbf{R},\mathbf{R}'}. \quad (39)$$

Now we can take the derivative:

$$\frac{\partial \langle \mathcal{H}_{\text{eff}} \rangle_0}{\partial h_\gamma} = 2 \sum_\delta m_\delta \frac{\partial m_\delta}{\partial h_\gamma} J_{\delta\delta} + 2 \sum_{\delta,\theta \neq \delta} m_\delta \frac{\partial m_\theta}{\partial h_\gamma} J_{\delta\theta}$$
$$= \boxed{2 \sum_{\delta,\theta} m_\delta \frac{\partial m_\delta}{\partial h_\gamma} J_{\delta\theta}.} \quad (40)$$

Our last term is:

$$(iii): \quad \langle \mathcal{H}_0 \rangle_0 = \left\langle -\sum_\gamma h_\gamma \sum_{\mathbf{R}}{}^{(\gamma)} \sigma_\mathbf{R} \right\rangle_0$$
$$= -\sum_\gamma h_\gamma m_\gamma N_\gamma$$
$$\implies \frac{\partial \langle \mathcal{H}_0 \rangle_0}{\partial h_\gamma} = \boxed{-m_\gamma N_\gamma - \sum_\delta h_\delta \frac{\partial m_\delta}{\partial h_\gamma} N_\delta.} \quad (41)$$

Putting everything together, we have:

$$0 = 2\sum_{\delta,\theta} m_\delta \frac{\partial m_\theta}{\partial h_\gamma} J_{\delta\theta} + \sum_\delta N_\delta \frac{\partial m_\delta}{\partial h_\gamma} h_\delta, \tag{42}$$

or, grouping terms by derivatives,

$$0 = \sum_\delta \frac{\partial m_\delta}{\partial h_\gamma} \left( 2\sum_\theta m_\theta J_{\delta\theta} + N_\delta h_\delta \right). \tag{43}$$

A solution to the above is:

$$h_\delta = -\frac{2}{N_\delta} \sum_\theta m_\theta J_{\delta\theta}. \tag{44}$$

We thus have a set of self-consistent equations:

$$m_\gamma = \tanh \beta \left( \mu - \frac{2}{N_\gamma} \sum_\delta m_\delta J_{\gamma\delta} \right), \tag{45}$$

subject to the constraint:

$$\bar{m} = \sum_\gamma m_\gamma x_\gamma, \tag{46}$$

where $x_\gamma = N_\gamma/N$. There may be multiple possible solutions to the above self-consistent equations, but the physically acceptable ones will minimize the free energy:

$$\begin{aligned} F(\bar{m}) &= -k_B T \log Q_{\text{est}}(\bar{m}) \\ &= -k_B T \sum_\gamma N_\gamma \log \left[ 2\cosh\left(\tanh^{-1} m_\gamma\right) \right] - \sum_{\gamma,\delta} J_{\gamma,\delta} m_\gamma m_\delta + \mu N \bar{m}. \end{aligned} \tag{47}$$

Equations 45, 46, and 47 constitute the mean field solution to the problem of modulated order at a fixed net composition. To apply it, we simply need to specify the sublattices and evaluate the coupling constants $J_{\gamma\delta}$. The sublattices comprising the triangular lattice are shown in Fig. 1. The self-consistent equations can be solved numerically to yield the equilibrium values for the sublattice magnetizations (and $\mu$.) We did so using Scipy's [10] root-finding routine "brentq" in the "optimize" package [11]. In Fig. 2 we plot our order parameter $\Delta m$, defined as the (magnitude of the) difference between the largest sublattice magnetization and the smallest sublattice magnetization, for a fixed composition of $c = 2/3$ ($\bar{m} = 1/3$.) We see that mean field theory predicts a (first-order) transition from a high temperature disordered state to a low-temperature state in which the sublattices attain net magnetizations different from $\bar{m}$. Specifically, one sublattice attains a negative magnetization, and the other two attain a positive magnetization. That is, one sublattice is occupied by predominantly $B$ atoms, while the other two are occupied by predominantly $A$ atoms.

## FRATZL-PENROSE-STYLE MEAN FIELD THEORY

Here we perform a mean field theory in the style of Fratzl and Penrose (which we refer to as MFT-2) [12], and compare it to the results of our mean field theory described in the previous section (which we refer to as MFT-1). The mean field free energy is given by:

$$F_{\text{MF}} = E - TS \tag{48}$$

$$= \sum_{\mathbf{R},\mathbf{R}'\neq\mathbf{R}} V_{\mathbf{R},\mathbf{R}'} \sigma_\mathbf{R} \sigma_{\mathbf{R}'} + k_B T \sum_\mathbf{R} \left( \frac{1+\sigma_\mathbf{R}}{2} \log\left(\frac{1+\sigma_\mathbf{R}}{2}\right) + \frac{1-\sigma_\mathbf{R}}{2} \log\left(\frac{1-\sigma_\mathbf{R}}{2}\right) \right). \tag{49}$$

Noting that $\sum_{\mathbf{R},\mathbf{R}'\neq\mathbf{R}} V_{\mathbf{R},\mathbf{R}'} \sigma_\mathbf{R} \sigma_{\mathbf{R}'} = \sum_{\mathbf{R},\mathbf{R}'} \sigma_\mathbf{R} (V_{\mathbf{R},\mathbf{R}'} - V_0 \delta_{\mathbf{R},\mathbf{R}'}) \sigma_{\mathbf{R}'}$, and setting the derivative with respect to the spin field to zero,

$$0 = \frac{\delta F_{\text{MF}}}{\delta \sigma_\mathbf{R}} = k_B T \frac{1}{2} \log\left(\frac{1+\sigma_\mathbf{R}}{1-\sigma_\mathbf{R}}\right) + 2\sum_{\mathbf{R}'} \sigma_{\mathbf{R}'} (V_{\mathbf{R},\mathbf{R}'} - V_0 \delta_{\mathbf{R},\mathbf{R}'}). \tag{50}$$

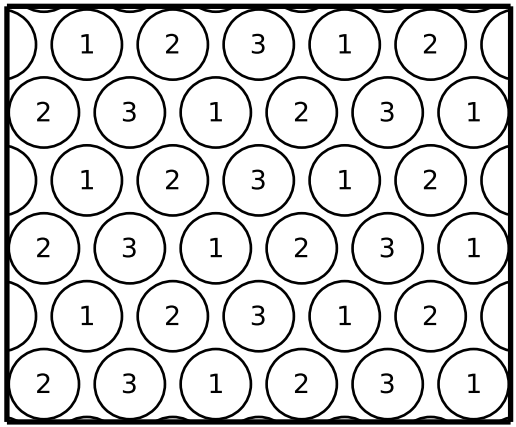

FIG. 1. The triangular lattice naturally decomposes into three intercalating sublattices, as labeled above.

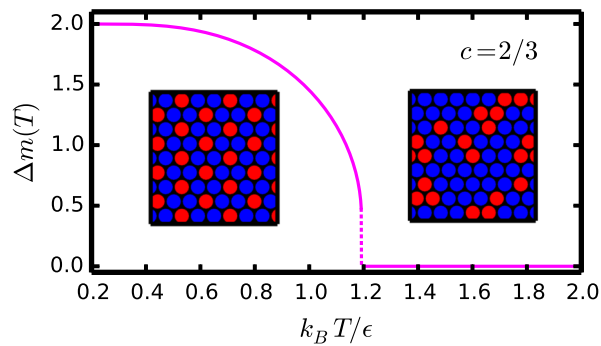

FIG. 2. Plot of the order parameter $\Delta m$ as a function of temperature, computed by solving the mean-field equations (45, 46, and 47) at a composition $c = 2/3$. The discontinuous jump in $\Delta m$ at $k_B T/\epsilon \approx 1.2$ (indicated by the dashed line) marks a first order phase transition from a disordered phase to a superlattice phase. A schematic superlattice configuration is shown on the left, while a schematic disordered configuration is shown on the right.

Some algebra shows that:

$$\frac{1}{2}\log\left(\frac{1+\sigma_{\mathbf{R}}}{1-\sigma_{\mathbf{R}}}\right) = -\tanh^{-1}(-\sigma_{\mathbf{R}}), \tag{51}$$

and so:

$$\sigma_{\mathbf{R}} = \tanh\left(-2\beta\sum_{\mathbf{R}'}\sigma_{\mathbf{R}'}(V_{\mathbf{R},\mathbf{R}'} - V_0\delta_{\mathbf{R},\mathbf{R}'})\right). \tag{52}$$

Now we write the spin field as:

$$\sigma_{\mathbf{R}} = \bar{m} + \Delta m \cos(\Delta\mathbf{G}\cdot\mathbf{R}), \tag{53}$$

where $\Delta\mathbf{G}$ is the wavevector on which superlattice ordering occurs. We will write equation 52 in terms of $\Delta m$, anticipating that at high temperatures in the absence of order, $\Delta m$ will be zero, whereas at low temperatures it will be nonzero. Plugging equation 53 into equation 52, writing cosine as a sum of complex exponentials, writing $V$ in terms of its Fourier coefficients, and simplifying, we find:

$$\bar{m} + \Delta m = \tanh\left[2\beta\left((V_0 - \tilde{V}_0)\bar{m} + \Delta m(V_0 - \tilde{V}_{\Delta\mathbf{G}})\right)\right]. \tag{54}$$

Unless $\bar{m} = 0$ equation 54 will not have $\Delta m = 0$ as a solution (as can be seen by taking the first term in a Taylor series expansion of the right hand side.) This is unphysical, because at high enough temperature there must be a

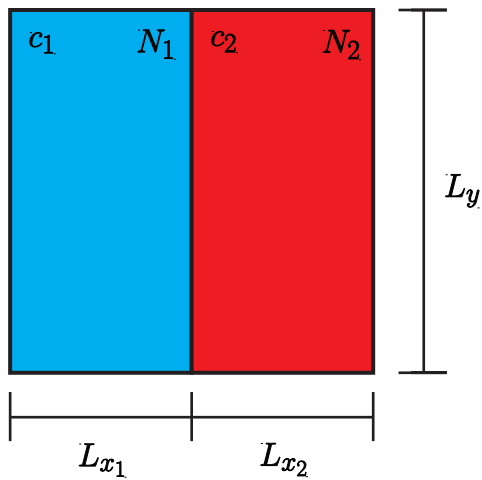

FIG. 3. Schematic of a phase-separated configuration.

solution $\Delta m = 0$ corresponding to a disordered phase. However, if we take $\bar{m} = 0$, we have:

$$\Delta m = \tanh\left[2\beta \Delta m (V_0 - \tilde{V}_{\Delta \mathbf{G}})\right]. \tag{55}$$

The transition will occur when $2\beta(V_0 - \tilde{V}_{\Delta \mathbf{G}}) = 1$, or:

$$k_B T_c = 2(V_0 - \tilde{V}_{\Delta \mathbf{G}}). \tag{56}$$

For our triangular lattice, $V_0 \approx 3.654$ and $\tilde{V}_{\Delta \mathbf{G}} = 3$. Therefore,

$$k_B T_c \approx 1.307. \tag{57}$$

This agrees quite well with our MFT-1 result:

$$k_B T_c \approx 1.308. \tag{58}$$

Hence, while MFT-2 cannot work for compositions other than $c = 1/2$, it actually quantitatively agrees with MFT-1 at $c = 1/2$.

## FREE ENERGY OF ELASTIC PHASE SEPARATION

Consider a mixture of two atomic species which have different natural bond lengths. This mixture is contained in a fixed two-dimensional area with a specified aspect ratio. Suppose that we separate this mixture into two domains of different composition. We assume that the interface between these domains is parallel to the $y$-axis. The domains, labeled 1 and 2, have compositions of $c_1$ and $c_2$, numbers of atoms $N_1$ and $N_2$, and widths of $L_{x_1}$ and $L_{x_2}$, respectively. Both domains have length $L_y$. This constraint on the dimensions of the domains causes elastic strain. A schematic of this setup is shown in Fig. 3. We define:

$$l_y = L_y/N_y, \tag{59}$$

where $N_y$ is the number of lattice sites along the $y$ direction, and:

$$n_i = N_i/N, \tag{60}$$

where $i = 1, 2$. Additionally, we have the following constraints:

$$n_1 + n_2 = 1 \tag{61a}$$
$$n_1 c_1 + n_2 c_2 = c. \tag{61b}$$

Now let us consider the free energy of phase separation. This is given by:

$$F = N_1 f(c_1, l_y) + N_2 f(c_2, l_y) + \text{(int.)}, \tag{62}$$

where $f$ is the free energy density of a homogeneous phase, and the final term is denotes energetic contributions with scale with the interfacial length, which we neglect in the thermodynamic limit. Noting that in an elastic medium there is a restoring force against changes in length, and that for small enough changes the associated energy is quadratic in the change, we write the free energy density as:

$$f(c_1, l_y) = f(c_1, \bar{l}_y(c_1)) + Y_y(l_y - \bar{l}_y)^2, \tag{63}$$

where $\bar{l}_y(c_1)$ is the unstrained value of $l_y$ for a homogeneous medium with composition $c_1$, and $Y_y$ is a Young's modulus for stretching or compressing the medium in the $y$ direction. In what follows, we shall drop the subscripts to make things cleaner, i.e. let $l_y = l$ and $Y_y = Y$. Now we make an assumption about the dependence of $\bar{l}$ on concentration, namely:

$$\bar{l}(c) = \bar{l}(0) + c\Delta l, \tag{64}$$

where $\Delta l \equiv \bar{l}(1) - \bar{l}(0)$. This assumption is known as Vegard's law [13] and we have observed in simulations that it holds for our microscopic elastic model.

Before proceeding to analyze the free energy, let us ask what the equilibrium value of $l$ is for the phase-separated system. This is determined by minimizing the free energy with respect to $l$:

$$\begin{aligned}
0 = \frac{\partial F/N}{\partial l} &= 2n_1 Y(l - \bar{l}(c_1)) + 2n_2 Y(l - \bar{l}(c_2)) \\
\Longrightarrow 0 &= (n_1 + n_2)l - (n_1 \bar{l}(c_1) + n_2 \bar{l}(c_2)) \\
&= l - n_1(\bar{l}(0) + c_1 \Delta l) - n_2(\bar{l}(0) + c_2 \Delta l) \\
&= l - \bar{l}(0) - (n_1 c_1 + n_2 c_2)\Delta l \\
&= l - \bar{l}(0) - c\Delta l \\
\Longrightarrow l &= \bar{l}(0) + c\Delta l,
\end{aligned} \tag{65}$$

where we have used equations 61a and 61b. This shows that the phase-separated system also obeys Vegard's Law. Let us use this fact to rewrite the free energy expression. Note that:

$$\begin{aligned}
l - \bar{l}(c_j) &= \bar{l}(0) + c\Delta l - (\bar{l}(0) + c_j \Delta l) \\
&= -\Delta c_j \Delta l,
\end{aligned}$$

where we have defined $\Delta c_j = c_j - c$. The free energy of phase separation can now be rewritten as:

$$F/N = n_1 \left( f(c_1) + Y \Delta l^2 \Delta c_1^2 \right) + n_2 \left( f(c_2) + Y \Delta l^2 \Delta c_2^2 \right). \tag{66}$$

We can eliminate $n_1$ and $n_2$ by making use of the following:

$$\begin{aligned}
c &= n_1 c_1 + n_2 c_2 \\
&= n_1(c + \Delta c_1) + n_2(c + \Delta c_2) \\
\Longrightarrow n_1 \Delta c_1 &= -n_2 \Delta c_2 \\
&= (n_1 - 1)\Delta c_2 \\
\Longrightarrow n_1 &= \frac{-\Delta c_2}{\Delta c_1 - \Delta c_2}.
\end{aligned} \tag{67a}$$

Similarly,

$$n_2 = \frac{\Delta c_1}{\Delta c_1 - \Delta c_2}. \tag{67b}$$

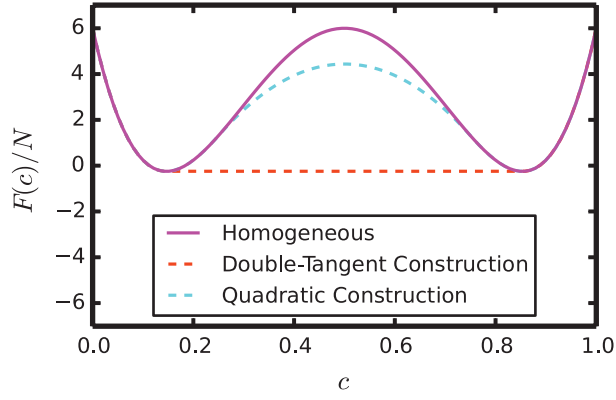

FIG. 4. Standard double-tangent construction and quadratic construction performed on a fictitious homogeneous free energy profile. For this fictitious quadratic construction, $\kappa_{\text{coex}}$ is taken to be $-50$.

Also, defining $\Delta f = f(c_2) - f(c_1)$, we write the free energy as:

$$
\begin{aligned}
F/N &= \frac{\Delta c_2}{\Delta c_2 - \Delta c_1} \left( f(c_1) + Y \Delta l^2 \Delta c_1^2 \right) \\
&\quad - \frac{\Delta c_1}{\Delta c_2 - \Delta c_1} \left( f(c_1) + \Delta f + Y \Delta l^2 \Delta c_2^2 \right) \\
&= f(c_1) - \frac{\Delta c_1}{\Delta c_2 - \Delta c_1} \Delta f + Y \Delta l^2 \frac{\Delta c_2 \Delta c_1^2 - \Delta c_1 \Delta c_2^2}{\Delta c_2 - \Delta c_1} \\
&= f(c_1) - \frac{\Delta c_1}{\Delta c_2 - \Delta c_1} \Delta f - Y \Delta l^2 \Delta c_1 \Delta c_2.
\end{aligned}
\tag{68}
$$

Equation 68 is our final result for the elastic free energy of coexistence. The first two terms constitute the usual double tangent construction - a straight line connecting the two coexisting compositions. The last term is a consequence of the macroscopic energy required to stretch the coexisting elastic domains so that they have the same length. It has a quadratic dependence on the overall composition, and hence we will not have a straight line connecting free energies at coexisting compositions, but a parabola.

### CONSTRUCTION OF THE COEXISTENCE FREE ENERGY

We identify the equilibrium compositions $c_j^*$ by evaluating $F_{\text{coex}}(c_1, c_2; c)$ for all values of $c_1$ and $c_2$ consistent with a given $c$, and selecting the pair that yields minimum free energy. Taking a graphical view of this solution, we could equivalently bring a parabola $f_0 - (\kappa_{\text{coex}}/2)(c - c_0)^2$ from below $f(c)$, varying parameters $f_0$ and $c_0$ until the two curves meet at a pair of points without crossing. All intervals of $c$ on which $\partial^2 f(c)/\partial c^2 < 2\kappa_{\text{coex}}$ are replaced in the QC by a parabola representing coexistence. The result of applying this procedure to a fictitious homogeneous free energy profile is shown in Fig. 4. The value of $\kappa_{\text{coex}}$ appropriate to our triangular lattice is $-16$; we derive this fact in the following section.

### MICROSCOPIC EXPRESSION FOR YOUNG'S MODULUS

It is possible to derive the Young's Modulus, and hence the energy of phase separation, microscopically. To do so, we decompose the triangular lattice into *plaquettes* - three-bond triangular loops. When phase separation occurs, two macroscopic domains with different natural sizes need to be stretched (or compressed) in order to have the same length. As long as we can think of the resultant strain in an individual domain as being uniform across plaquettes, we can obtain the energy associated with phase separation simply by evaluating the energy of an individual plaquette. Consider a triangle with side lengths $a_1$ (lying along the $x$-axis) and $a_2 = a_3$ (with components in both $x$ and $y$.) When unstrained, these lengths are all equal to the lattice constant $a$. We denote the height of this triangle as $l_y = \sqrt{3}a/2$.

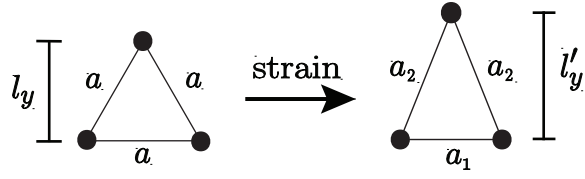

FIG. 5. Illustration of a triangular plaquette undergoing a deformation applied in the $y$ (vertical) direction.

When a uniform strain is applied to the elastic domain in, say, the $y$ direction, the bonds in the plaquettes will adopt new lengths in order to minimize the elastic energy. This situation is illustrated in Fig. 5. Given a lattice mismatch $\Delta$ between the two coexisting elastic domains, the new height $l'_y$ is given by:

$$l'_y = \frac{\sqrt{3}}{2}(a + \Delta). \tag{69}$$

Our task now is to find the strained values of $a_1$, $a_2$, and $a_3$. We do so by minimizing the energy of the strained plaquette. The energies of the individual bonds are given by:

$$\epsilon_1 = \frac{K}{2}(a_1 - a)^2 \tag{70}$$

$$\epsilon_2 = \epsilon_3 = \frac{K}{2}(a_2(a_1) - a)^2, \tag{71}$$

where, by geometry, we have written $a_2$ as a function of $a_1$:

$$a_2(a_1) = \sqrt{\left(\frac{a_1}{2}\right)^2 + \frac{3}{4}(a + \Delta)^2}. \tag{72}$$

The total energy of the plaquette is then given by:

$$\epsilon_{\text{tot}} = \epsilon_1 + \epsilon_2 + \epsilon_3 \tag{73}$$

$$= \frac{K}{2}\left[(a_1 - a)^2 + 2(a_2(a_1) - a)^2\right]. \tag{74}$$

To obtain the minimum energy and associated bond lengths, we minimize $\epsilon_{\text{tot}}$ with respect to $a_1$:

$$0 = \frac{\partial \epsilon_{\text{tot}}}{\partial a_1} = K(a_1 - a) + 2K(a_2(a_1) - a)\frac{\partial a_2}{\partial a_1} \tag{75}$$

$$\implies 0 = a_1\left(1 + \frac{1}{2}\left(\frac{a_2 - a}{a_2}\right)\right) - a. \tag{76}$$

Rearranging, we obtain a self-consistent equation for $a_1$:

$$a_1 = \frac{a}{\frac{3}{2} - \frac{a}{2a_2(a_1)}} \tag{77}$$

Assuming that $a_1$ is not too different from $a$, we can linearize the self-consistent equation about $a_1 = a$. Expanding the resulting expression to first order in $\Delta$ (assumed to be small,) we find:

$$a_1 \approx a - \frac{\Delta}{3}. \tag{78}$$

Plugging this expression into the equation for $a_2$ and again expanding to first order in $\Delta$, we have:

$$a_2 \approx a + \frac{2}{3}\Delta. \tag{79}$$

The plaquette energy is therefore:

$$\epsilon_{\text{tot}} \approx \frac{K}{2}\Delta^2. \tag{80}$$

Multiplying by $N$, the total number of plaquettes, gives the total energy of phase separation. To extract the Young's modulus, compare this expression for the energy of phase separation with Eq. 63. This yields:

$$Y = \frac{2}{3}K, \tag{81}$$

which agrees with the value of $Y$ we infer by applying mechanical deformation in a simulated, zero-temperature elastic domain and measuring the resulting elastic energy as a function of deformation. In the energy units used in the main text $\epsilon = K\Delta^2/8$, we have $Y\Delta l^2 = 16$.

## ENERGETIC STABILITY OF COEXISTING PHASES

Here we assess the stability of coexisting elastic phases of different compositions with respect to the creation of interfaces. To do so, we explicitly evaluate the energy of configurations with macroscopic spatial modulations of varying wavelength $\lambda$. If this energy decreases with increasing $\lambda$, then coexistence is energetically stable. The stability of coexistence will turn out to depend on the internal structure of the phases.

### Form of the spatial modulations in Fourier space

We first write the spatially modulated composition field in Fourier space. In real space, a composition field $\sigma(\mathbf{r})$ which consists of a series of domains of alternating composition separated by sharp interfaces ("stripes") can be written as:

$$\sigma(\mathbf{r}) = \frac{1}{2}\left(\sigma_1(\mathbf{r}) + \sigma_2(\mathbf{r})\right) + s(\mathbf{r})\left(\sigma_1(\mathbf{r}) - \sigma_2(\mathbf{r})\right), \tag{82}$$

where $s(\mathbf{r})$ is a square wave whose direction of modulation is along an as-yet-unspecified axis, and where $\sigma_\mu(\mathbf{r})$ ($\mu = 1, 2$) are the (potentially spatially-dependent) composition fields of the two different macroscopic phases. In Fourier space, this field can be written as:

$$\tilde{\sigma}(\mathbf{q}) = \frac{1}{2}\left(\tilde{\sigma}_1(\mathbf{q})) + \tilde{\sigma}_2(\mathbf{q})\right) + \frac{1}{N}\sum_{\mathbf{q}'}\tilde{s}(\mathbf{q}-\mathbf{q}')\left(\tilde{\sigma}_1(\mathbf{q}') - \tilde{\sigma}_2(\mathbf{q}')\right), \tag{83}$$

where we have used the fact that the Fourier transform of a product of two functions in real space is their convolution in Fourier space. Before we can go any further, we need to determine the forms of $\tilde{\sigma}_\mu(\mathbf{q})$. In real space, since we allow for the possibility that the individual macroscopic phases can have periodic microscopic order, we write the composition field as:

$$\sigma_\mu(\mathbf{r}) = \sum_{\mathbf{b}_\mu} c_{\mathbf{b}_\mu} e^{i\mathbf{b}_\mu \cdot \mathbf{r}}, \tag{84}$$

where $\mathbf{b}_\mu$ are reciprocal lattice vectors. In Fourier space, this becomes:

$$\tilde{\sigma}_\mu(\mathbf{q}) = \sum_{\mathbf{b}_\mu} N c_{\mathbf{b}_\mu} \delta(\mathbf{q}, \mathbf{b}_\mu). \tag{85}$$

This expression can be derived straightforwardly:

$$\tilde{\sigma}_\mu(\mathbf{q}) = \sum_{\mathbf{r}} \sigma_\mu(\mathbf{r}) e^{-i\mathbf{q}\cdot\mathbf{r}}$$

$$= \sum_{\mathbf{r}} \left( \sum_{\mathbf{b}_\mu} c_{\mathbf{b}_\mu} e^{i\mathbf{b}_\mu \cdot \mathbf{r}} \right) e^{-i\mathbf{q}\cdot\mathbf{r}}$$

$$= \sum_{\mathbf{b}_\mu} c_{\mathbf{b}_\mu} \left( \sum_{\mathbf{r}} e^{i(\mathbf{b}_\mu - \mathbf{q})\cdot\mathbf{r}} \right)$$

$$= \sum_{\mathbf{b}_\mu} N c_{\mathbf{b}_\mu} \delta(\mathbf{q}, \mathbf{b}_\mu).$$

The Fourier coefficients $c_{\mathbf{b}_\mu}$ can be determined by examining the real space composition field. For a phase with spatially uniform composition, $\sigma_\mu(\mathbf{r}) = \bar{\sigma}$, we have:

$$c_{\mathbf{b}_\mu} = \bar{\sigma}\delta(\mathbf{b}_\mu, 0), \tag{86}$$

while for a superlattice phase, where $\sigma_\mu(\mathbf{r}) = -\frac{1}{3} + \frac{4}{3}\cos(\Delta\mathbf{G}\cdot\mathbf{r})$ (or its negative): [14]

$$c_{\mathbf{b}_\mu} = -\frac{1}{3}\delta(\mathbf{b}_\mu, 0) + \frac{2}{3}\delta(\mathbf{b}_\mu, \Delta\mathbf{G}) + \frac{2}{3}\delta(\mathbf{b}_\mu, -\Delta\mathbf{G}). \tag{87}$$

Now, for both spatially uniform and superlattice phases, $c_{\mathbf{b}_\mu} = c_{-\mathbf{b}_\mu}$, so we can write:

$$\tilde{\sigma}_\mu(\mathbf{q}) = \sideset{}{'}\sum_{\mathbf{b}_\mu} N c_{\mathbf{b}_\mu} \left( \delta(\mathbf{q}, \mathbf{b}_\mu) + \delta(\mathbf{q}, -\mathbf{b}_\mu) \right), \tag{88}$$

where the prime symbol indicates that the sum is to be taken over the positive half space, and implies a factor of $1/2$ when $\mathbf{b}_\mu = 0$. Furthermore we separate the $\mathbf{b}_\mu = 0$ component from the sum and write:

$$\tilde{\sigma}_\mu(\mathbf{q}) = N c_{0,\mu} \delta(\mathbf{q}, 0) + \sideset{}{''}\sum_{\mathbf{b}_\mu} N c_{\mathbf{b}_\mu} \left( \delta(\mathbf{q}, \mathbf{b}_\mu) + \delta(\mathbf{q}, -\mathbf{b}_\mu) \right), \tag{89}$$

where the extra prime symbol indicates that the $\mathbf{b}_\mu = 0$ term is excluded from the sum. The Fourier coefficient $c_{0,\mu}$ is equal to $\bar{\sigma}_\mu$, the net composition of phase $\mu$. With this expression for the single-phase, Fourier-space composition

fields in hand, we can write the overall Fourier-space composition field as:

$$\tilde{\sigma}(\mathbf{q}) = N\left(\frac{\bar{\sigma}_1+\bar{\sigma}_2}{2}\right)\delta(\mathbf{q},0) + \frac{1}{2}\sum_{\mu=1,2}\sum_{\mathbf{b}_\mu}{}'' Nc_{\mathbf{b}_\mu}\left(\delta(\mathbf{q},\mathbf{b}_\mu)+\delta(\mathbf{q},-\mathbf{b}_\mu)\right)$$

$$+\frac{1}{N}\sum_{\mathbf{q}'}\tilde{s}(\mathbf{q}-\mathbf{q}')\left(N\bar{\sigma}_1\delta(\mathbf{q}',0)+\sum_{\mathbf{b}_1}{}'' Nc_{\mathbf{b}_1}\left(\delta(\mathbf{q}',\mathbf{b}_1)+\delta(\mathbf{q}',-\mathbf{b}_1)\right)\right.$$

$$\left.-N\bar{\sigma}_2\delta(\mathbf{q}',0)-\sum_{\mathbf{b}_2}{}'' Nc_{\mathbf{b}_2}\left(\delta(\mathbf{q}',\mathbf{b}_1 2+\delta(\mathbf{q}',-\mathbf{b}_2))\right)\right)$$

$$= N\left(\frac{\bar{\sigma}_1+\bar{\sigma}_2}{2}\right)\delta(\mathbf{q},0)+\frac{1}{2}\sum_{\mu=1,2}\sum_{\mathbf{b}_\mu}{}'' Nc_{\mathbf{b}_\mu}\left(\delta(\mathbf{q},\mathbf{b}_\mu)+\delta(\mathbf{q},-\mathbf{b}_\mu)\right)$$

$$+\tilde{s}(\mathbf{q})\bar{\sigma}_1-\tilde{s}(\mathbf{q})\bar{\sigma}_2+\sum_{\mathbf{b}_1}{}'' c_{\mathbf{b}_1}\left(\tilde{s}(\mathbf{q}-\mathbf{b}_1)+\tilde{s}(\mathbf{q}+\mathbf{b}_1)\right)$$

$$-\sum_{\mathbf{b}_2}{}'' c_{\mathbf{b}_2}\left(\tilde{s}(\mathbf{q}-\mathbf{b}_2)+\tilde{s}(\mathbf{q}+\mathbf{b}_2)\right)$$

$$= N\left(\frac{\bar{\sigma}_1+\bar{\sigma}_2}{2}\right)\delta(\mathbf{q},0)+(\bar{\sigma}_1-\bar{\sigma}_2)\tilde{s}(\mathbf{q})$$

$$+\sum_{\mathbf{b}_1}{}'' c_{\mathbf{b}_1}\left(\frac{N}{2}\delta(\mathbf{q},\mathbf{b}_1)+\frac{N}{2}\delta(\mathbf{q},-\mathbf{b}_1)+\tilde{s}(\mathbf{q}-\mathbf{b}_1)+\tilde{s}(\mathbf{q}+\mathbf{b}_1)\right)$$

$$+\sum_{\mathbf{b}_2}{}'' c_{\mathbf{b}_2}\left(\frac{N}{2}\delta(\mathbf{q},\mathbf{b}_2)+\frac{N}{2}\delta(\mathbf{q},-\mathbf{b}_2)-\tilde{s}(\mathbf{q}-\mathbf{b}_2)-\tilde{s}(\mathbf{q}+\mathbf{b}_2)\right). \tag{90}$$

The next step before evaluating the energy is to obtain an expression for $\tilde{s}(\mathbf{q})$, the square wave in Fourier space.

### Fourier-space representation of the square wave

Writing the square wave in Fourier space requires some care. Let us start by choosing a basis in which to evaluate it. The standard choice is:

$$\mathbf{a}_1 = (1,0) \tag{91a}$$

$$\mathbf{a}_2 = \left(\frac{1}{2},\frac{\sqrt{3}}{2}\right). \tag{91b}$$

The corresponding reciprocal lattice vectors are:

$$\mathbf{b}_1 = \left(2\pi,-\frac{2\pi}{\sqrt{3}}\right) \tag{92a}$$

$$\mathbf{b}_2 = \left(0,\frac{4\pi}{\sqrt{3}}\right). \tag{92b}$$

If the stripes have normal vectors parallel to the y direction $\hat{\mathbf{y}}$, then $\mathbf{b}_2$ is parallel to those normal vectors, which makes life easier. Restricting ourselves to the first Brillouin zone of reciprocal space, we write:

$$\mathbf{q} = \frac{k_1}{N_1}\mathbf{b}_1+\frac{k_2}{N_2}\mathbf{b}_2 \tag{93}$$

$$\mathbf{r} = n_1\mathbf{a}_1+n_2\mathbf{a}_2. \tag{94}$$

So, $s(\mathbf{r})$ is indexed by $(n_1,n_2)$ and $\tilde{s}(\mathbf{q})$ is indexed by $(k_1,k_2)$. For stripes along $\hat{\mathbf{y}}$, we have:

$$s(n_1,n_2)-s(n_1,n_2-1) = \begin{cases} -1, & n_2 = 2\lambda, 4\lambda,\ldots \\ 1, & n_2 = \lambda, 3\lambda,\ldots \\ 0, & \text{otherwise,} \end{cases} \tag{95}$$

where $\lambda$ is the width of the stripe. Making the definitions:

$$\tilde{s}(k_1, k_2) = \sum_{n_1, n_2} s(n_1, n_2) e^{2\pi i \left(\frac{n_1 k_1}{N_1} + \frac{n_2 k_2}{N_2}\right)} \tag{96}$$

$$s(n_1, n_2) = \frac{1}{N_1 N_2} \sum_{k_1, k_2} \tilde{s}(k_1, k_2) e^{-2\pi i \left(\frac{n_1 k_1}{N_1} + \frac{n_2 k_2}{N_2}\right)}, \tag{97}$$

we can write:

$$s(n_1, n_2) - s(n_1, n_2 - 1) = \frac{1}{N_1 N_2} \sum_{k_1, k_2} \tilde{s}(k_1, k_2) \left( e^{-2\pi i \left(\frac{n_1 k_1}{N_1} + \frac{n_2 k_2}{N_2}\right)} - e^{-2\pi i \left(\frac{n_1 k_1}{N_1} + \frac{(n_2-1) k_2}{N_2}\right)} \right)$$

$$= \frac{1}{N_1 N_2} \sum_{k_1, k_2} \tilde{s}(k_1, k_2) e^{-i\mathbf{q}\cdot\mathbf{r}} (1 - e^{2\pi i k_2 / N_2}). \tag{98}$$

At the same time, we can write:

$$s(n_1, n_2) - s(n_1, n_2 - 1) = \delta(n_2, \lambda) - \delta(n_2, 2\lambda) + \delta(n_2, 3\lambda) - \cdots$$

$$= \frac{1}{N_2} \sum_{k_2} \left( e^{-2\pi i k_2 (n_2 - \lambda)/N_2} - e^{-2\pi i k_2 (n_2 - 2\lambda)/N_2} + \cdots \right)$$

$$\times \frac{1}{N_1} \sum_{k_1} e^{-2\pi i k_1 n_1 / N_1} N_1 \delta(k_1, 0). \tag{99}$$

Comparing equations 98 and 99, we have:

$$\tilde{s}(k_1, k_2)(1 - e^{2\pi i k_2/N_2}) = N_1 \delta(k_1, 0) \left( e^{2\pi i k_2 \lambda / N_2} - e^{2\pi i k_2 (2\lambda)/N_2} + \cdots \right)$$

$$\boxed{\implies \tilde{s}(k_1, k_2) = (1 - e^{2\pi i k_2/N_2})^{-1} \delta(k_1, 0) \frac{N_1 N_2}{\lambda} \sum_{j=\pm 1, \pm 3, \ldots}^{\lambda} -\delta\left(k_2, \frac{N_2 j}{2\lambda}\right).}$$

If, on the other hand, the stripes are parallel to the x direction $\hat{\mathbf{x}}$, then the previous choice of basis is awkward because neither reciprocal lattice vector is parallel to $\hat{\mathbf{x}}$. (It is very convenient to describe a spatial modulation when one of the reciprocal lattice vectors points in the direction of that modulation.) Instead we pick a new basis:

$$\mathbf{a}_1 = \left(\frac{1}{2}, \frac{\sqrt{3}}{2}\right) \tag{100a}$$

$$\mathbf{a}_2 = \left(0, \sqrt{3}\right), \tag{100b}$$

in which the corresponding reciprocal lattice vectors are:

$$\mathbf{b}_1 = (4\pi, 0) \tag{101a}$$

$$\mathbf{b}_2 = \left(-2\pi, \frac{2\pi}{\sqrt{3}}\right). \tag{101b}$$

(Note that $\mathbf{a}_2$ is not normalized.) Now $\mathbf{b}_1$ is parallel to $\hat{\mathbf{x}}$, which is the direction of modulation. Again, we write:

$$\mathbf{q} = \frac{k_1}{N_1} \mathbf{b}_1 + \frac{k_2}{N_2} \mathbf{b}_2 \tag{102}$$

$$\mathbf{r} = n_1 \mathbf{a}_1 + n_2 \mathbf{a}_2. \tag{103}$$

Now, our square wave will satisfy:

$$s(n_1, n_2) - s(n_1 - 1, n_2) = \begin{cases} -1, & n_1 = 2\lambda, 4\lambda, \ldots \\ 1, & n_1 = \lambda, 3\lambda, \ldots \\ 0, & \text{otherwise}, \end{cases} \tag{104}$$

Pulling the same computational tricks as before, we arrive at:

$$\tilde{s}(k_1, k_2) = (1 - e^{2\pi i k_1/N_1})^{-1} \delta(k_2, 0) \frac{N_1 N_2}{\lambda} \sum_{j=\pm 1, \pm 3, \ldots}^{\lambda} -\delta\left(k_1, \frac{N_1 j}{2\lambda}\right). \quad (105)$$

This is essentially the same form we had before, but we have switched the role of $\mathbf{b}_1$ and $\mathbf{b}_2$.

### Evaluating stripe energies

Now we evaluate the energies for stripes with different kinds of microscopic structure.

*Case 1: No microscopic structure*

In the case of no microscopic structure in either phase, we have $c_{\mathbf{b}_\mu} = \bar{\sigma}_\mu \delta(\mathbf{b}_\mu, 0)$. Hence the composition field is given by:

$$\tilde{\sigma}(\mathbf{q}) = N\left(\frac{\bar{\sigma}_1 + \bar{\sigma}_2}{2}\right) \delta(\mathbf{q}, 0) + (\bar{\sigma}_1 - \bar{\sigma}_2) \tilde{s}(\mathbf{q}). \quad (106)$$

The energy of this field is:

$$E = \frac{1}{2N} \sum_{\mathbf{q}} |\tilde{\sigma}(\mathbf{q})|^2 \tilde{V}(\mathbf{q})$$

$$= \frac{(\bar{\sigma}_1 - \bar{\sigma}_2)^2}{2N} \sum_{\mathbf{q}} |\tilde{s}(\mathbf{q})|^2 \tilde{V}(\mathbf{q})$$

$$= \frac{\Delta\bar{\sigma}^2}{2} \sum_{k_1, k_2} \frac{\delta(k_2, 0)}{2 - 2\cos(2\pi k_1/N_1)} \frac{1}{\lambda^2} \sum_{j=\pm 1, \ldots}^{\lambda} \delta\left(k_1, \frac{N_1 j}{2\lambda}\right) \tilde{V}(k_1, k_2)$$

$$= \frac{\Delta\bar{\sigma}^2}{2\lambda^2} \sum_{k_1} \frac{1}{2 - 2\cos(2\pi k_1/N_1)} \sum_{j=\pm 1, \ldots}^{\lambda} \delta\left(k_1, \frac{N_1 j}{2\lambda}\right) \tilde{V}(k_1, 0)$$

$$= \boxed{\frac{\Delta\bar{\sigma}^2}{2\lambda^2} \sum_{j=\pm 1, \ldots}^{\lambda} \frac{\tilde{V}\left(\frac{N_1 j}{2\lambda}, 0\right)}{2 - 2\cos(\pi j/\lambda)}}.$$

(In the above we made the definition $\Delta\bar{\sigma} = \bar{\sigma}_1 - \bar{\sigma}_2$.) If we view $\tilde{V}$ as a function of $q_x$, $q_y$ rather than $k_1$, $k_2$, then since $\mathbf{q} \cdot \hat{\mathbf{x}} = 2\pi j/\lambda$, we have:

$$E(\lambda) = \frac{\Delta\bar{\sigma}^2}{2\lambda^2} \sum_{j=\pm 1, \ldots}^{\lambda} \frac{\tilde{V}\left(\frac{2\pi j}{\lambda}, 0\right)}{2 - 2\cos(\pi j/\lambda)}. \quad (107)$$

*Case 2: One spatially uniform phase, one superlattice phase*

In the case of one spatially uniform phase and one superlattice phase, we have:

$$c_{\mathbf{b}_1} = \delta(\mathbf{b}_1, 0)$$
$$c_{\mathbf{b}_2} = \frac{1}{3}\delta(\mathbf{b}_2, 0) - \frac{2}{3}\delta(\mathbf{b}_2, \Delta\mathbf{G}) - \frac{2}{3}\delta(\mathbf{b}_2, -\Delta\mathbf{G}).$$

The composition field is then given by:

$$\tilde{\sigma}(\mathbf{q}) = N\left(\frac{\bar{\sigma}_1 + \bar{\sigma}_2}{2}\delta(\mathbf{q},0)\right) + \Delta\bar{\sigma}^2|\tilde{s}(\mathbf{q})|^2$$
$$-\frac{2}{3}\left(\frac{N}{2}\delta(\mathbf{q},\Delta\mathbf{G}) + \frac{N}{2}\delta(\mathbf{q},-\Delta\mathbf{G}) - \tilde{s}(\mathbf{q}-\Delta\mathbf{G}) - \tilde{s}(\mathbf{q}+\Delta\mathbf{G})\right). \quad (108)$$

Taking the square modulus of the field (noting that many of the possible terms vanish due to the factor $\delta(k_2,0)$ contained in $\tilde{s}$,) we have:

$$|\tilde{\sigma}(\mathbf{q})|^2 = N^2\left(\frac{\bar{\sigma}_1 + \bar{\sigma}_2}{2}\right)^2\delta(\mathbf{q},0) + \Delta\bar{\sigma}^2|\tilde{s}(\mathbf{q})|^2$$
$$+\frac{4}{9}\left(N^2\delta(\mathbf{q},\Delta\mathbf{G}) + N^2\delta(\mathbf{q},-\Delta\mathbf{G}) + |\tilde{s}(\mathbf{q}-\Delta\mathbf{G})|^2 + |\tilde{s}(\mathbf{q}+\Delta\mathbf{G})|^2\right). \quad (109)$$

The energy is:

$$E(\lambda) = \frac{1}{2N}\sum_{\mathbf{q}}|\tilde{\sigma}(\mathbf{q})|^2\tilde{V}(\mathbf{q})$$
$$= \frac{\Delta\bar{\sigma}^2}{2N}\sum_{\mathbf{q}}|\tilde{s}(\mathbf{q})|^2\tilde{V}(\mathbf{q}) + \frac{1}{9}N\left(\tilde{V}(\Delta\mathbf{G}) + \tilde{V}(-\Delta\mathbf{G})\right)$$
$$+\frac{2}{9}\frac{1}{N}\sum_{\mathbf{q}}\left(|\tilde{s}(\mathbf{q}-\Delta\mathbf{G})|^2 + |\tilde{s}(\mathbf{q}+\Delta\mathbf{G})|^2\right)\tilde{V}(\mathbf{q}).$$

Writing:

$$\Delta\mathbf{G} = \frac{\alpha_1}{N_1}\mathbf{b}_1 + \frac{\alpha_2}{N_2}\mathbf{b}_2, \quad (110)$$

we have:

$$\tilde{s}(\mathbf{q}-\Delta\mathbf{G}) = \tilde{s}(k_1-\alpha_1, k_2-\alpha_2)$$
$$= \left(1 - e^{2\pi i(k_1-\alpha_1)/N_1}\right)^{-1}\delta(k_2,\alpha_2)\frac{N_1}{\lambda}\sum_{j=\pm 1,\ldots}^{\lambda}\delta\left(k_1, \alpha_1 + \frac{N_1 j}{2\lambda}\right).$$

Thus we can write the energy as:

$$E(\lambda) = \frac{\Delta\bar{\sigma}^2}{2\lambda^2}\sum_{j=\pm 1,\ldots}^{\lambda}\frac{\tilde{V}\left(\frac{N_1 j}{2\lambda},0\right)}{2-2\cos(\pi j/\lambda)} + \frac{2}{9}N\tilde{V}(\Delta\mathbf{G})$$
$$+\frac{2}{9\lambda^2}\left(\sum_{j=\pm 1,\ldots}^{\lambda}\frac{\tilde{V}\left(\alpha_1 + \frac{N_1 j}{2\lambda},\alpha_2\right)}{2-2\cos(\pi j/\lambda)} + \sum_{j=\pm 1,\ldots}^{\lambda}\frac{\tilde{V}\left(-\alpha_1 + \frac{N_1 j}{2\lambda},-\alpha_2\right)}{2-2\cos(\pi j/\lambda)}\right).$$

Collecting terms,

$$\boxed{\begin{aligned}E(\lambda) = \frac{1}{9}N\tilde{V}(\Delta\mathbf{G}) + \frac{1}{2\lambda^2}\sum_{j=\pm 1,\ldots}^{\lambda}&\left(\frac{\Delta\bar{\sigma}^2\tilde{V}\left(\frac{N_1 j}{2\lambda},0\right)}{2-2\cos(\pi j/\lambda)}\right.\\ &\left.+\frac{4}{9}\frac{\left(\tilde{V}\left(\alpha_1 + \frac{N_1 j}{2\lambda},\alpha_2\right) + \tilde{V}\left(-\alpha_1 + \frac{N_1 j}{2\lambda},-\alpha_2\right)\right)}{2-2\cos(\pi j/\lambda)}\right).\end{aligned}} \quad (111)$$

Again, we could express $\tilde{V}$ in terms of $q_x$ and $q_y$ instead of $k_1$ and $k_2$. To do so, note that:

$$3\Delta\mathbf{G} = 2\pi(1, -\sqrt{3})$$
$$= -3\mathbf{b}_2 - \mathbf{b}_1$$
$$\implies \Delta\mathbf{G} = -\frac{1}{3}\mathbf{b}_1 - \mathbf{b}_2$$
$$= -\frac{N_1/3}{N_1}\mathbf{b}_1 - \frac{N_2}{N_2}\mathbf{b}_2.$$

So,

$$\alpha_1 = -N_1/3$$
$$\alpha_2 = -N_2.$$

Then it is just a matter of computing $\mathbf{q} \cdot \hat{\mathbf{x}}$ and $\mathbf{q} \cdot \hat{\mathbf{y}}$ from the definition of $\mathbf{q}$.

*Case 3: Two superlattice phases in register*

In the case of two superlattice phases, we have:

$$c_{\mathbf{b}_1} = \frac{1}{3}\delta(\mathbf{b}_2, 0) - \frac{2}{3}\delta(\mathbf{b}_2, \Delta\mathbf{G}) - \frac{2}{3}\delta(\mathbf{b}_2, -\Delta\mathbf{G})$$
$$c_{\mathbf{b}_2} = -\frac{1}{3}\delta(\mathbf{b}_2, 0) + \frac{2}{3}\delta(\mathbf{b}_2, \Delta\mathbf{G}) + \frac{2}{3}\delta(\mathbf{b}_2, -\Delta\mathbf{G}).$$

In this section we shall assume that the two phases are "in register," i.e. that the interface between the superlattices is not offset by some vector. In this case, the composition field is:

$$\tilde{\sigma}(\mathbf{q}) = \Delta\bar{\sigma}\tilde{s}(\mathbf{q}) - \frac{4}{3}\left(\tilde{s}(\mathbf{q} - \Delta\mathbf{G}) + \tilde{s}(\mathbf{q} + \Delta\mathbf{G})\right), \tag{112}$$

so,

$$|\tilde{\sigma}(\mathbf{q})|^2 = \Delta\bar{\sigma}^2|\tilde{s}(\mathbf{q})|^2 + \frac{16}{9}\left(|\tilde{s}(\mathbf{q} - \Delta\mathbf{G})|^2 + |\tilde{s}(\mathbf{q} + \Delta\mathbf{G})|^2\right). \tag{113}$$

The energy is:

$$E(\lambda) = \frac{1}{2N} \sum_{\mathbf{q}} |\tilde{\sigma}(\mathbf{q})|^2 \tilde{V}(\mathbf{q})$$
$$= \frac{\Delta\bar{\sigma}^2}{2N} \sum_{\mathbf{q}} |\tilde{s}(\mathbf{q})|^2 \tilde{V}(\mathbf{q}) + \frac{8}{9N} \sum_{\mathbf{q}} \left(\tilde{s}(\mathbf{q} - \Delta\mathbf{G})|^2 + |\tilde{s}(\mathbf{q} + \Delta\mathbf{G})|^2\right) \tilde{V}(\mathbf{q})$$
$$= \boxed{\frac{1}{2\lambda^2} \sum_{j=\pm 1,\ldots}^{\lambda} \frac{\Delta\bar{\sigma}^2 \tilde{V}\left(\frac{N_1 j}{2\lambda}, 0\right) + \frac{16}{9}\left(\tilde{V}\left(\alpha_1 + \frac{N_1 j}{2\lambda}, \alpha_2\right) + \tilde{V}\left(-\alpha_1 + \frac{N_1 j}{2\lambda}, -\alpha_2\right)\right)}{2 - 2\cos(\pi j/\lambda)}}. \tag{114}$$

*Case 4: Two superlattice phases out of register*

In this section we shall allow for the possibility that the two phases are out of register by some vector $\mathbf{a}$. The composition fields of the individual phases are given by:

$$\sigma_1(\mathbf{r}) = \frac{1}{3} - \frac{4}{3}\cos(\mathbf{r} \cdot \Delta\mathbf{G})$$
$$\sigma_2(\mathbf{r}) = -\frac{1}{3} + \frac{4}{3}\cos((\mathbf{r} + \mathbf{a}) \cdot \Delta\mathbf{G}),$$

and so:

$$c_{\mathbf{b}_1} = \frac{1}{3}\delta(\mathbf{b}_2, 0) - \frac{2}{3}\delta(\mathbf{b}_2, \Delta\mathbf{G}) - \frac{2}{3}\delta(\mathbf{b}_2, -\Delta\mathbf{G})$$
$$c_{\mathbf{b}_2} = -\frac{1}{3}\delta(\mathbf{b}_2, 0) + \frac{2}{3}e^{i\Delta\mathbf{G}\cdot\mathbf{a}}\delta(\mathbf{b}_2, \Delta\mathbf{G}) + \frac{2}{3}e^{-i\Delta\mathbf{G}\cdot\mathbf{a}}\delta(\mathbf{b}_2, -\Delta\mathbf{G}).$$

In this case, the composition field is:

$$\tilde{\sigma}(\mathbf{q}) = \Delta\bar{\sigma}\tilde{s}(\mathbf{q}) + \frac{N}{3}(e^{i\Delta\mathbf{G}\cdot\mathbf{a}} - 1)(\delta(\mathbf{q}, \Delta\mathbf{G}) + \delta(\mathbf{q}, -\Delta\mathbf{G}))$$
$$- \frac{2}{3}(e^{i\Delta\mathbf{G}\cdot\mathbf{a}} + 1)(\tilde{s}(\mathbf{q} - \Delta\mathbf{G}) + \tilde{s}(\mathbf{q} + \Delta\mathbf{G})),$$

so,

$$|\tilde{\sigma}(\mathbf{q})|^2 = \Delta\bar{\sigma}^2|\tilde{s}(\mathbf{q})|^2 + \frac{N^2}{9}(2 - 2\cos(\Delta\mathbf{G}\cdot\mathbf{a}))(\delta(\mathbf{q}, \Delta\mathbf{G}) + \delta(\mathbf{q}, -\Delta\mathbf{G}))$$
$$+ \frac{4}{9}(2 + 2\cos(\Delta\mathbf{G}\cdot\mathbf{a}))\left(|\tilde{s}(\mathbf{q} - \Delta\mathbf{G})|^2 + |\tilde{s}(\mathbf{q} + \Delta\mathbf{G})|^2\right),$$

and hence the energy is:

$$E(\lambda) = \frac{1}{2N}\sum_{\mathbf{q}} |\tilde{\sigma}(\mathbf{q})|^2 \tilde{V}(\mathbf{q})$$
$$= \frac{\Delta\bar{\sigma}^2}{2N}\sum_{\mathbf{q}} |\tilde{s}(\mathbf{q})|^2 \tilde{V}(\mathbf{q}) + \frac{N}{18}(2 - 2\cos(\Delta\mathbf{G}\cdot\mathbf{a}))\left(\tilde{V}(\Delta\mathbf{G}) + \tilde{V}(-\Delta\mathbf{G})\right)$$
$$+ \frac{2}{9N}(2 + 2\cos(\Delta\mathbf{G}\cdot\mathbf{a}))\sum_{\mathbf{q}}\left(|\tilde{s}(\mathbf{q} - \Delta\mathbf{G})|^2 + |\tilde{s}(\mathbf{q} + \Delta\mathbf{G})|^2\right)\tilde{V}(\mathbf{q}).$$

Plugging in $\tilde{s}$ and simplifying we have, finally,

$$\boxed{\begin{aligned}E(\lambda) = &\frac{1}{9}N(2 - 2\cos(\Delta\mathbf{G}\cdot\mathbf{a}))\tilde{V}(\Delta\mathbf{G}) + \frac{1}{2\lambda^2}\sum_{j=\pm 1,\ldots}^{\lambda}\left(\frac{\Delta\bar{\sigma}^2\tilde{V}\left(\frac{N_1 j}{2\lambda}, 0\right)}{2 - 2\cos(\pi j/\lambda)}\right.\\ &\left.+\frac{4}{9}\frac{(2 + 2\cos(\Delta\mathbf{G}\cdot\mathbf{a}))\left(\tilde{V}\left(\alpha_1 + \frac{N_1 j}{2\lambda}, \alpha_2\right) + \tilde{V}\left(-\alpha_1 + \frac{N_1 j}{2\lambda}, -\alpha_2\right)\right)}{2 - 2\cos(\pi j/\lambda)}\right).\end{aligned}} \qquad (115)$$

*A single superlattice domain*

We need the energy of a single bulk superlattice in order to compare stripe energies. To do so, note that for a superlattice the composition field is given by:

$$\tilde{\sigma}(\mathbf{q}) = -\frac{1}{3}N\delta(\mathbf{q}, 0) + \frac{2}{3}N\delta(\mathbf{q}, \Delta\mathbf{q}) + \frac{2}{3}N\delta(\mathbf{q}, -\Delta\mathbf{q}), \qquad (116)$$

so the energy is:

$$E_{\text{super}} = \frac{1}{2N}\sum_{\mathbf{q}} |\tilde{\sigma}(\mathbf{q})|^2 \tilde{V}(\mathbf{q})$$
$$= \frac{2}{9}N\left(\tilde{V}(\Delta\mathbf{G}) + \tilde{V}(-\Delta\mathbf{G})\right)$$
$$= \frac{4}{9}N\tilde{V}(\Delta\mathbf{G}).$$

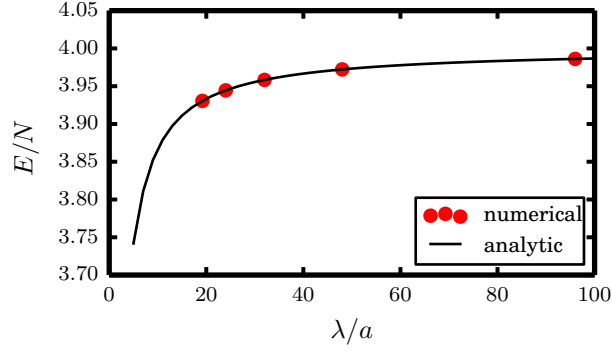

FIG. 6. Energy associated with unstructured stripes (with uniform compositions.)

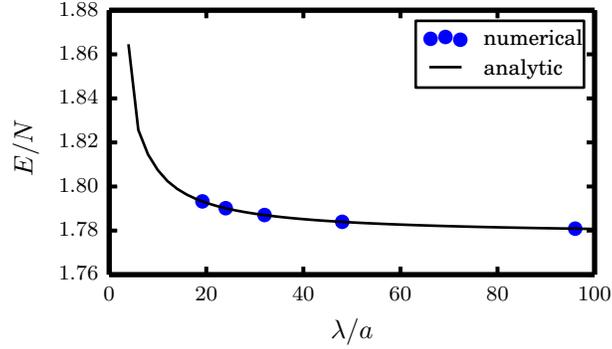

FIG. 7. Energy of superlattice/unstructured stripes.

**Numerical evaluation of energies**

Now we can numerically evaluate the energies of different stripe configurations. Since $\tilde{V}(\Delta \mathbf{G}) = 6$, the energy, per atom, of a single superlattice domain is:

$$E_{\text{super}}/N = \frac{4}{9}\tilde{V}(\Delta \mathbf{G}) = \frac{8}{3}. \tag{117}$$

This allows us to evaluate the energy associated with coexistence between phases present in equal proportions. Numerically evaluating the sums in the energy expressions, setting $\lambda = 10000$, we have:

$$E_{\text{unstruct+unstruct}}/N \approx 4.01492$$
$$E_{\text{unstruct+super}}/N - \frac{1}{2}E_{\text{super}}/N \approx 0.446148$$
$$E_{\text{super+super}} - E_{\text{super}}/N \approx 0.446280.$$

These are the energy penalties associated with coexistence between two unstructured phases, an unstructured and a superlattice phase, and two (in-register) superlattice phases, respectively. (Note that we have subtracted off the energies of the bulk phases.) For all $\lambda$, our theoretical results agree essentially exactly with the results of energy minimization of our simulated system. Plots comparing these results are shown in Figs. 6, 7 and 8. Coexistence between unstructured phases is seen to be unstable (energy increases with $\lambda$,) but coexistence between an unstructured phase and a superlattice phase, or two superlattice phases, is energetically stable (energy decreases with $\lambda$.)

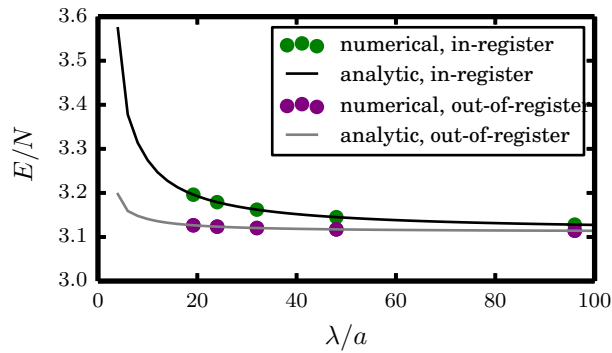

FIG. 8. Energy of superlattice/superlattice stripes. The black line and green dots are for an in-register interface; the gray line and purple dots are for an out-of-register interface.

## ENERGETICS OF IMPURITY CONFIGURATIONS

The effective potential arises from the microscopic strain required to accommodate atoms of different sizes in the same lattice. Here we plot configurations of bonds, along with their associated strain, for a few "impurity" configurations, in which one or two atoms of type $B$ have been substituted in an otherwise perfect $A$ crystal. These zero-temperature configurations were generated by performing energy minimization via gradient descent.

Fig. 9 shows a single $B$ atom surrounded by $A$ atoms. The "inner shell" of bonds emanating from the $B$ atom have natural length $l_{AB}$, but in this minimum-energy configuration they do not achieve this length. They are instead stretched out by $A$ atoms, which prefer a longer bond length. Meanwhile, the "second shell" of bonds connecting those of the inner shell are compressed relative to their natural length of $l_{AA}$.

Fig. 10 shows the result of making two such impurities nearest neighbors. The bond connecting them has a natural length $l_{BB} < l_{AB}$. Evidently, minimizing the total energy means that this bond is far from its natural length. Lattice strain is highly localized in this bond and its immediate neighbors to the left and right. The resulting large energy contribution disfavors this configuration relative to a configuration in which the $B$ atoms are very far apart. This is a consequence of the fact that a pair of $B$ atoms prefers a bond distance $l_{BB}$ that is smaller than either $l_{AA}$ or $l_{AB}$. This preference is strongly frustrated when adjacent $B$ defects are surrounded by an excess of $A$ atoms, making this arrangement especially costly. The two defects are more readily accommodated on next-nearest neighbor sites. In this configuration, the energy is lower than that of infinite separation (see Fig. 11). Similarly, two $A$ defects in a $B$-rich environment can mitigate the cost of local expansion by sharing an intervening $B$ neighbor. Inspection of the configuration reveals that the "second shell" bond shared by the $B$ atoms bears a significant amount of strain, although much less than the $B-B$ bond in the nearest-neighbor configuration. Additional strain appears to be shared more equitably (compared to the nearest-neighbor configuration) among neighboring bonds, resulting in a lower total energy. Despite the more even distribution of strain, compression of the shared second shell bond appears to be necessary for the energetic favorability of this configuration; if we fix bond's length to the value it would have if the impurity were isolated, then the next-nearest neighbor configuration becomes energetically unfavorable compared to two isolated impurities. This reasoning allows us to microscopically rationalize the anisotropy of our effective potential.

## ASYMPTOTIC SCALING OF EFFECTIVE POTENTIAL

In the low-$q$ limit, Taylor expansion of our effective potential yields leading order terms $\sim 4 - a(\hat{\mathbf{q}})q^2$, where $a$ is a function of the direction of $\mathbf{q}$ but not its magnitude. The inverse Fourier transform is anisotropic, but scales as $\sim 1/R^4$ [15]. We confirmed this scaling by plotting the numerical inverse Fourier transform of $\tilde{V}_{\mathbf{q}}$ along different directions. Figs. 12 and 13 show the magnitude of the effective potential along two different lattice directions (one along one of the six principal lattice directions, here $\hat{\mathbf{a}} = (1, 0)$ and another along a vector halfway between two principal lattice directions, $\hat{\mathbf{a}} = (\sqrt{3}/2, 1/2)$. In both cases, fits to the tail of the energy yield a power law with exponent $-4$, consistent with our theoretical prediction.

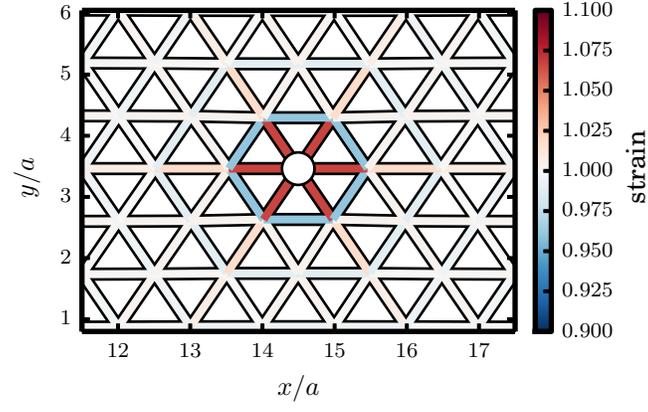

FIG. 9. Strain (defined as $|\mathbf{r_R} - \mathbf{r_{R+a\hat{\alpha}}}|/l(\sigma_\mathbf{R},\sigma_{\mathbf{R}+a\hat{\alpha}})$) of bonds due to the presence of a single $B$ defect surrounded by $A$ atoms. The defect is denoted with a white circle.

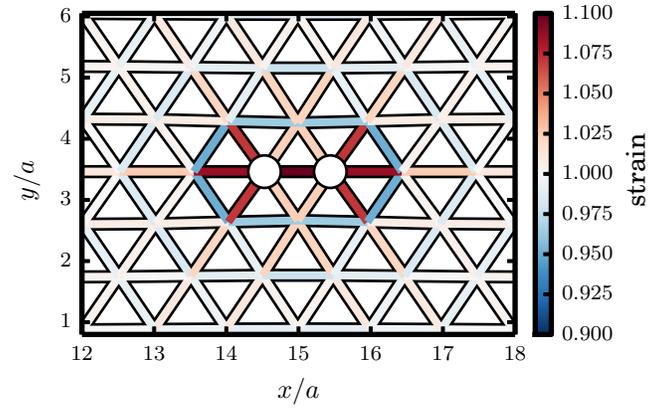

FIG. 10. Strain due to the presence of two nearest-neighbor $B$ defects surrounded by $A$ atoms.

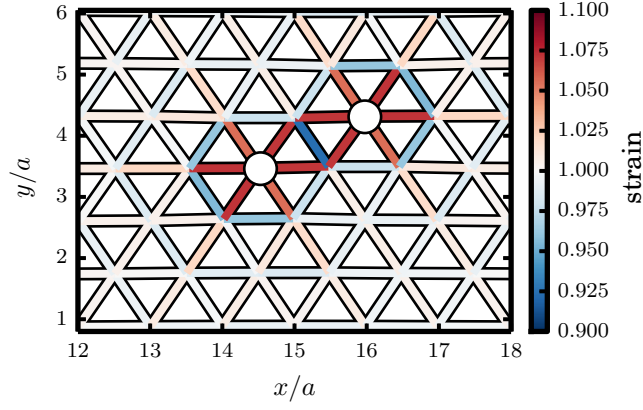

FIG. 11. Strain due to the presence of two next-nearest-neighbor $B$ defects surrounded by $A$ atoms.

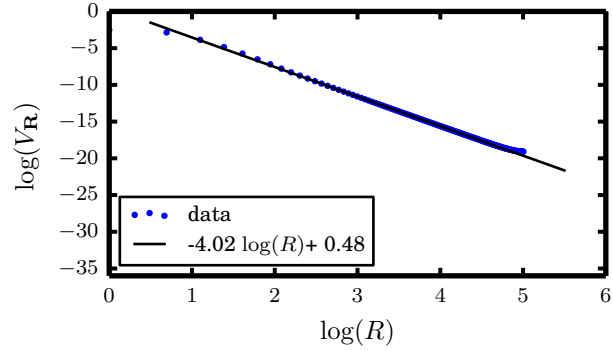

FIG. 12. Effective potential along the direction defined by the unit vector $\hat{\mathbf{a}} = (1, 0)$. Data was obtained from the numerical inverse Fourier transform of $\tilde{V}_{\mathbf{q}}$ on a 300 by 300 unit cell triangular lattice. The solid black line is a fit to the data, excluding the first 10 points and last 90 points (which are sensitive to the finite size of the lattice.)

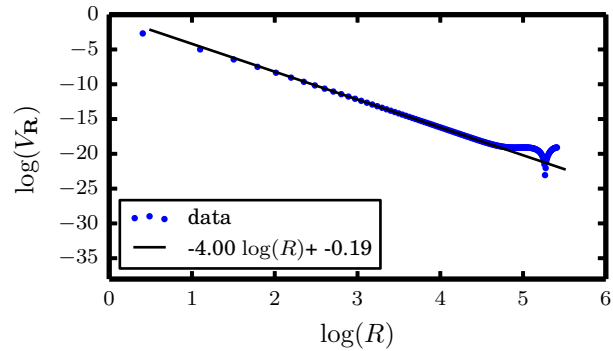

FIG. 13. Effective potential along the direction defined by the unit vector $\hat{\mathbf{a}} = (\sqrt{3}/2, 1/2)$. Data was obtained from the numerical inverse Fourier transform of $\tilde{V}_{\mathbf{q}}$ on a 300 by 300 unit cell triangular lattice. The solid black line is a fit to the data, excluding the first 10 points and last 90 points (which are sensitive to the finite size of the lattice.)

# EFFECTIVE POTENTIAL IN 3D

Following the same procedure we used for the two-dimensional triangular lattice, we obtained an effective potential for the three-dimensional FCC lattice. The bond vectors are:

$$\hat{\boldsymbol{\alpha}}_1 = \begin{pmatrix} 1/\sqrt{2} \\ 1/\sqrt{2} \\ 0 \end{pmatrix} \tag{118a}$$

$$\hat{\boldsymbol{\alpha}}_2 = \begin{pmatrix} 1/\sqrt{2} \\ 0 \\ 1/\sqrt{2} \end{pmatrix} \tag{118b}$$

$$\hat{\boldsymbol{\alpha}}_3 = \begin{pmatrix} 0 \\ 1/\sqrt{2} \\ 1/\sqrt{2} \end{pmatrix} \tag{118c}$$

$$\hat{\boldsymbol{\alpha}}_4 = \begin{pmatrix} 0 \\ -1/\sqrt{2} \\ 1/\sqrt{2} \end{pmatrix} \tag{118d}$$

$$\hat{\boldsymbol{\alpha}}_5 = \begin{pmatrix} -1/\sqrt{2} \\ 0 \\ 1/\sqrt{2} \end{pmatrix} \tag{118e}$$

$$\hat{\boldsymbol{\alpha}}_6 = \begin{pmatrix} -1/\sqrt{2} \\ 1/\sqrt{2} \\ 0 \end{pmatrix}, \tag{118f}$$

(as well as their additive inverses) and the effective potential is:

$$\tilde{V}_{\mathbf{q}} = A(\mathbf{q})/B(\mathbf{q}), \tag{119}$$

where now $\mathbf{q} = (q_x, q_y, q_z)$ and:

$$A(\mathbf{q}) = 4\left(8\cos\left(\frac{q_z}{\sqrt{2}}\right)\cos\left(\frac{q_x-q_y}{2\sqrt{2}}\right)\left(-41\cos\left(\frac{q_x}{\sqrt{2}}\right)\cos\left(\frac{q_y}{\sqrt{2}}\right) + \cos\left(\sqrt{2}q_x\right)\left(\cos\left(\sqrt{2}q_y\right) + 3\right) + 3\cos\left(\sqrt{2}q_y\right) + 64\right)\right.$$
$$\times \cos\left(\frac{q_x+q_y}{2\sqrt{2}}\right) + 8\cos\left(\frac{3q_z}{\sqrt{2}}\right)\cos\left(\frac{q_x-q_y}{2\sqrt{2}}\right)\left(\cos\left(\frac{q_x}{\sqrt{2}}\right)\cos\left(\frac{q_y}{\sqrt{2}}\right) + 1\right)\cos\left(\frac{q_x+q_y}{2\sqrt{2}}\right) + \cos\left(\sqrt{2}q_z\right)$$
$$\times \left(\cos\left(\frac{q_x-3q_y}{\sqrt{2}}\right) - 34\cos\left(\frac{q_x-q_y}{\sqrt{2}}\right) + 6\cos\left(\sqrt{2}(q_x-q_y)\right) + \cos\left(\frac{3q_x-q_y}{\sqrt{2}}\right) - 34\cos\left(\frac{q_x+q_y}{\sqrt{2}}\right)\right.$$
$$+ 6\cos\left(\sqrt{2}(q_x+q_y)\right) + \cos\left(\frac{3q_x+q_y}{\sqrt{2}}\right) + \cos\left(\frac{q_x+3q_y}{\sqrt{2}}\right) - 4\cos\left(\sqrt{2}q_x\right) - 4\cos\left(\sqrt{2}q_y\right) - 36\right) + 3\cos\left(\frac{q_x-3q_y}{\sqrt{2}}\right)$$
$$+ 90\cos\left(\frac{q_x-q_y}{\sqrt{2}}\right) - 2\cos\left(\sqrt{2}(q_x-q_y)\right) + 3\cos\left(\frac{3q_x-q_y}{\sqrt{2}}\right) + 90\cos\left(\frac{q_x+q_y}{\sqrt{2}}\right) - 2\cos\left(\sqrt{2}(q_x+q_y)\right)$$
$$+ 3\cos\left(\frac{3q_x+q_y}{\sqrt{2}}\right) + 3\cos\left(\frac{q_x+3q_y}{\sqrt{2}}\right) - 36\cos\left(\sqrt{2}q_x\right) - 36\cos\left(\sqrt{2}q_y\right) - 276\right), \tag{120}$$

$$B(\mathbf{q}) = 2\cos\left(\frac{q_z}{\sqrt{2}}\right)\left(\cos\left(\frac{q_x}{\sqrt{2}}\right)\left(22 - 8\cos\left(\sqrt{2}q_y\right)\right) + \cos\left(\frac{q_y}{\sqrt{2}}\right)\left(-8\cos\left(\sqrt{2}q_x\right) + 2\cos\left(\sqrt{2}q_y\right) + 21\right)\right.$$
$$+ \cos\left(\frac{3q_x}{\sqrt{2}}\right)\right) + 2\cos\left(\frac{3q_z}{\sqrt{2}}\right)\left(\cos\left(\frac{q_x}{\sqrt{2}}\right) + \cos\left(\frac{q_y}{\sqrt{2}}\right)\right) + 4\cos\left(\sqrt{2}q_z\right)\left(-4\cos\left(\frac{q_x}{\sqrt{2}}\right)\cos\left(\frac{q_y}{\sqrt{2}}\right)\right.$$
$$+ \cos\left(\sqrt{2}q_x\right)\cos\left(\sqrt{2}q_y\right) - 3\right) + \cos\left(\frac{q_x-3q_y}{\sqrt{2}}\right) + 22\cos\left(\frac{q_x-q_y}{\sqrt{2}}\right)$$
$$+ \cos\left(\frac{3q_x-q_y}{\sqrt{2}}\right) + 22\cos\left(\frac{q_x+q_y}{\sqrt{2}}\right) + \cos\left(\frac{3q_x+q_y}{\sqrt{2}}\right) + \cos\left(\frac{q_x+3q_y}{\sqrt{2}}\right) - 12\cos\left(\sqrt{2}q_x\right) - 12\cos\left(\sqrt{2}q_y\right) - 64. \tag{121}$$

A plot is shown in Fig. 14. The potential discourages spatial modulations of the composition which do not lie along the $\hat{\mathbf{x}}$, $\hat{\mathbf{y}}$, or $\hat{\mathbf{z}}$ directions. However, it is equally permissive of modulations of any wavelength along these principal directions, owing to the fact that the potential is essentially flat along these directions. This vast degeneracy could in principle be broken by the introduction of an additional spatially short-ranged interaction or, alternatively, the addition of next-nearest-neighbor springs [16]. We also computed the effective potential for a simple cubic lattice, on which next-nearest-neighbor springs are required to provide resistance to shear [17]. The nearest-neighbor bond vectors in that case are:

$$\hat{\boldsymbol{\alpha}}_1 = \begin{pmatrix} 1 \\ 0 \\ 0 \end{pmatrix} \tag{122a}$$

$$\hat{\boldsymbol{\alpha}}_2 = \begin{pmatrix} 0 \\ 1 \\ 0 \end{pmatrix} \tag{122b}$$

$$\hat{\boldsymbol{\alpha}}_3 = \begin{pmatrix} 0 \\ 0 \\ 1 \end{pmatrix}, \tag{122c}$$

as well as additive inverses, and the next-nearest-neighbor bond vectors $\{\hat{\boldsymbol{\beta}}_j\}$ are identical to the nearest-neighbor bond vectors on the FCC lattice. The associated effective potential is given by:

$$\tilde{V}_{\mathbf{q}} = 4 \left( C(\mathbf{q}) - \frac{D_1(\mathbf{q}) + D_2(\mathbf{q}) + D_3(\mathbf{q})}{E(\mathbf{q})} \right), \tag{123}$$

where:

$$C(\mathbf{q}) = \cos(q_x - q_y) + \cos(q_x + q_y) + \cos(q_x - q_z) \\ + \cos(q_x + q_z) + \cos(q_x) + \cos(q_y - q_z) + \cos(q_y + q_z) + \cos(q_y) + \cos(q_z) + 9 \quad (124a)$$

$$D_1(\mathbf{q}) = 2\sin^2(q_z)\left(\sqrt{2}\cos(q_x) + \sqrt{2}\cos(q_y) + 1\right) \\ \times \left(4\sin^2(q_x)\left(\sqrt{2}\cos(q_y) + \sqrt{2}\cos(q_z) + 1\right)(\cos(q_y)(\cos(q_x) + \cos(q_z) + 1) + \sin^2(q_y) - 3) \right. \\ + 4\sin^2(q_y)\left(\sqrt{2}\cos(q_x) + \sqrt{2}\cos(q_z) + 1\right) \times (\cos(q_x)(\cos(q_y) + \cos(q_z) + 1) + \sin^2(q_x) - 3) \\ + \left(\sqrt{2}\cos(q_x) + \sqrt{2}\cos(q_y) + 1\right)((2\cos(q_y)(\cos(q_x) + \cos(q_z) + 1) - 6) \\ \left. \times (2\cos(q_x)(\cos(q_y) + \cos(q_z) + 1) - 6) - 4\sin^2(q_x)\sin^2(q_y))\right) \quad (124b)$$

$$D_2(\mathbf{q}) = 2\sin^2(q_y)\left(\sqrt{2}\cos(q_x) + \sqrt{2}\cos(q_z) + 1\right) \\ \times \left(4\sin^2(q_x)\left(\sqrt{2}\cos(q_y) + \sqrt{2}\cos(q_z) + 1\right)(\cos(q_z)(\cos(q_x) + \cos(q_y) + 1) + \sin^2(q_z) - 3) \right. \\ + 4\sin^2(q_z)\left(\sqrt{2}\cos(q_x) + \sqrt{2}\cos(q_y) + 1\right)(\cos(q_x)(\cos(q_y) + \cos(q_z) + 1) + \sin^2(q_x) - 3) \\ + \left(\sqrt{2}\cos(q_x) + \sqrt{2}\cos(q_z) + 1\right)((2\cos(q_z)(\cos(q_x) + \cos(q_y) + 1) - 6) \\ \left. \times (2\cos(q_x)(\cos(q_y) + \cos(q_z) + 1) - 6) - 4\sin^2(q_x)\sin^2(q_z))\right) \quad (124c)$$

$$D_3(\mathbf{q}) = 2\sin^2(q_x)\left(\sqrt{2}\cos(q_y) + \sqrt{2}\cos(q_z) + 1\right) \\ \times \left(4\sin^2(q_y)\left(\sqrt{2}\cos(q_x) + \sqrt{2}\cos(q_z) + 1\right)(\cos(q_z)(\cos(q_x) + \cos(q_y) + 1) + \sin^2(q_z) - 3) \right. \\ + 4\sin^2(q_z)\left(\sqrt{2}\cos(q_x) + \sqrt{2}\cos(q_y) + 1\right)(\cos(q_y)(\cos(q_x) + \cos(q_z) + 1) + \sin^2(q_y) - 3) \\ + \left(\sqrt{2}\cos(q_y) + \sqrt{2}\cos(q_z) + 1\right)((2\cos(q_z)(\cos(q_x) + \cos(q_y) + 1) - 6) \\ \left. \times (2\cos(q_y)(\cos(q_x) + \cos(q_z) + 1) - 6) - 4\sin^2(q_y)\sin^2(q_z))\right) \quad (124d)$$

$$E(\mathbf{q}) = 8\sin^2(q_y)\sin^2(q_z)\left(\cos(q_x)(\cos(q_y) + \cos(q_z) + 1) + \sin^2(q_x) - 3\right) \\ + 8\sin^2(q_x)\sin^2(q_z)\left(\cos(q_y)(\cos(q_x) + \cos(q_z) + 1) + \sin^2(q_y) - 3\right) \\ + (6 - 2\cos(q_z)(\cos(q_x) + \cos(q_y) + 1)) \\ \times \left((2\cos(q_y)(\cos(q_x) + \cos(q_z) + 1) - 6)(2\cos(q_x)(\cos(q_y) + \cos(q_z) + 1) - 6) - 4\sin^2(q_x)\sin^2(q_y)\right) \quad (124e)$$

A plot is shown in 15. Unlike the FCC lattice, the minima here occur at isolated, nonzero, points. This suggests that finite-wavelength modulated order is energetically favorable on the simple cubic lattice, which we confirm with an exploratory MC simulation. In Fig. 16 we show a configuration taken from one such MC simulation of atoms interacting via the effective potential Eq. 123. The simulation consisted of $10^5$ MC sweeps on a $10 \times 10 \times 10$ simple cubic lattice at a temperature $T = 0.5$ and fixed composition $c = 1/2$. Starting from a random initial configuration, the system quickly reorganized into what we surmise is a modulated order ground state for this value of $c$, consisting of single-species columns of atoms arranged in an alternating "checkerboard" pattern.

## MONTE CARLO PHASE BEHAVIOR

Previous work by D. Landau [18] showed that a triangular lattice spin model with nearest-neighbor attraction and next-nearest neighbor repulsion (henceforth referred to as the "Landau model,") produces rich phase behavior. Of particular interest is the fact that mean field theory fails to capture some major features of the phase diagram. While mean field theory accounts for the low temperature phases and their coexistence scenarios, it entirely misses a higher-temperature region which exhibits critical lines with 3-state Potts universality and which end in tricritical points, as well as a line of Kosterlitz-Thouless critical points. We have determined a very similar phase diagram for our elastic model. In this section, we outline how we established various features of the phase behavior.

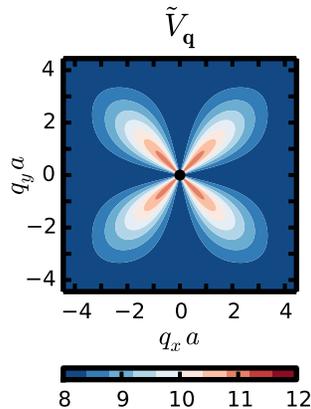

FIG. 14. Cross section of $\tilde{V}_\mathbf{q}$ for the FCC lattice at $q_z = 0$. The color indicates the magnitude of $\tilde{V}_\mathbf{q}$. The black dot in the center indicates the fact that the potential is zero at $\mathbf{q} = 0$.

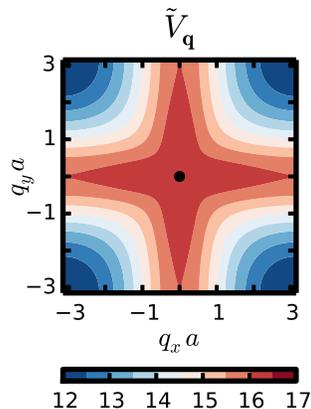

FIG. 15. Cross section of $\tilde{V}_\mathbf{q}$ for the simple cubic lattice at $q_z = 0$. The color indicates the magnitude of $\tilde{V}_\mathbf{q}$. The black dot in the center indicates the fact that the potential is zero at $\mathbf{q} = 0$.

### 3-State Potts Critical Lines

As previously mentioned, Landau found that for compositions $c \neq 1/2$ there are critical lines separating a high-temperature disordered phase and lower-temperature modulated-order ("superlattice") phases. Symmetry arguments have been advanced which state that in models like Landau's such critical lines should have three-state Potts universality [19]. This phase behavior is also very similar to that of the triangular lattice antiferromagnet in an external field. Transfer matrix results for that model are also consistent with three-state Potts universality [20]. We used finite-size scaling methods in order to locate critical lines in our model and to show that these critical lines have three-state Potts universality.

A common method for locating critical points involves computation of the Binder cumulant [21]:

$$U_N = 1 - \frac{\langle \Delta m^4 \rangle}{3 \langle \Delta m^2 \rangle^2}, \tag{125}$$

where $\Delta m$ is the order parameter. One can identify the critical point as the point of intersection of Binder cumulants for different system sizes. We used this technique to locate the critical lines. For several different fixed compositions, we computed $U_N$ as a function of temperature for several different system sizes. This is illustrated for a composition $c = 1/3$ in Fig. 17.

We also used finite-size scaling to confirm that the critical points located with the Binder cumulants have the critical

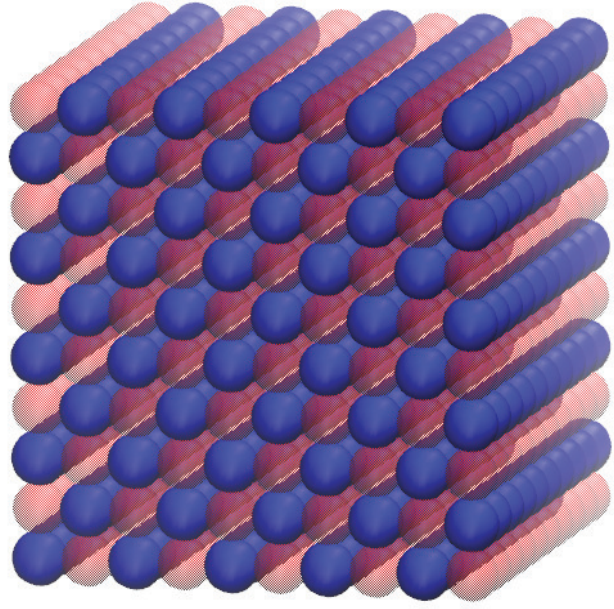

FIG. 16. Configuration taken from an MC simulation of atoms interacting according to the pair potential in Eq. 123. Columns of single-color (-species) stripes alternate in a checkerboard pattern. Red atoms (species B) are made translucent for clarity.

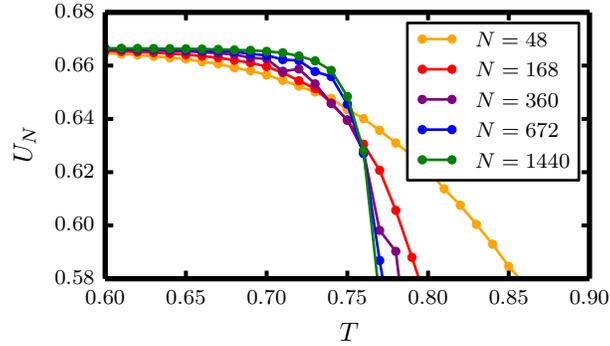

FIG. 17. Binder cumulants as a function of temperature for different system sizes at a fixed composition $c = 1/3$. Averages were taken over $10^6$ MC sweeps. The transition temperature was identified as $T_c \approx 0.76$.

exponents associated with the 3-state Potts model. Near a critical point, thermodynamic quantities like the specific heat $C = \left(\langle E^2 \rangle - \langle E \rangle^2\right)/(Nk_BT^2)$, the susceptibility $\chi = \langle \Delta m^2 \rangle$, and the correlation length $\xi$ exhibit power-law dependencies on the reduced temperature $t = (T - T_c)/T_c$ [22]:

$$C \propto |t|^{-\alpha} \tag{126a}$$
$$\Delta m \propto |t|^{\beta} \tag{126b}$$
$$\chi \propto |t|^{-\gamma} \tag{126c}$$
$$\xi \propto |t|^{-\nu}. \tag{126d}$$

For the two-dimensional three-state Potts model, $\alpha = 1/3$, $\beta = 1/9$, $\gamma = 13/9$, and $\nu = 5/6$ [18]. Extracting these exponents directly from measurements of these thermodynamic quantities in simulations is complicated by finite-size effects. Rather than trying to estimate these quantities directly, we can take advantage of finite-size scaling theory to confirm the values of the critical exponents. Specifically, one can write finite-size scaling expressions for

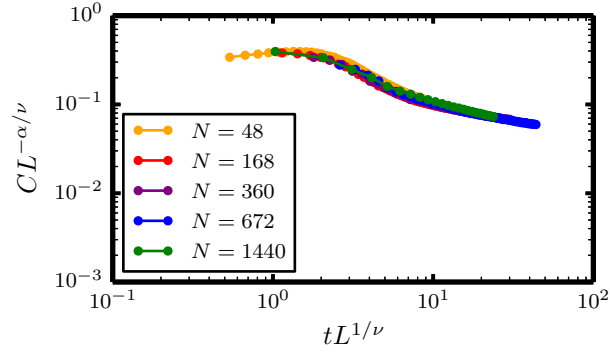

FIG. 18. Heat capacity for different system sizes at $c = 1/3$. Averages were taken over $10^6$ MC sweeps.

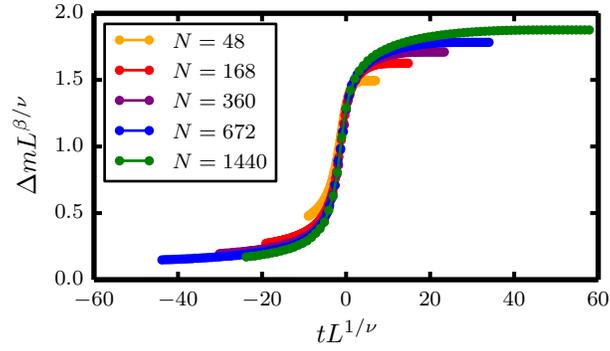

FIG. 19. Order parameter for different system sizes at $c = 1/3$. Averages were taken over $10^6$ MC sweeps.

thermodynamic quantities [18]:

$$C = L^{\alpha/\nu} f_1(x) \tag{127a}$$

$$\Delta m = L^{-\beta/\nu} f_2(x) \tag{127b}$$

$$\chi = L^{\gamma/\nu} f_3(x), \tag{127c}$$

where $x = tL^{1/\nu}$ is a scaled temperature and the functions $f_1$, $f_2$, and $f_3$ depend only on $x$. Therefore, the quantities $L^{-\alpha/\nu}C$, $L^{\beta/\nu}\Delta m$, and $L^{-\gamma/\nu}\chi$ as functions of $x$ are independent of system size. When plotted, data for different system sizes should therefore lie on top of each other close to $t = 0$. (This is known as "data collapse.") We made such plots to confirm that our critical lines indeed obey 3-state Potts universality. Examples are shown in Figs. 18, 19, and 20.

### Kosterlitz-Thouless Transition Line

At a composition $c = 1/2$ the Landau model exhibits a line of Kosterlitz-Thouless transitions, with upper critical temperature $T_c^{\text{upper}}$ and lower critical temperature $T_c^{\text{lower}}$. Near such a transition, the correlation length diverges exponentially [18]:

$$\xi = \xi_0 \exp(at^{-1/2}). \tag{128}$$

This leads to finite-size scaling expressions:

$$\Delta m = L^{-b} g_1(L^{-1} \exp(at^{-1/2})) \tag{129a}$$

$$\chi = L^c g_2(L^{-1} \exp(at^{-1/2})), \tag{129b}$$

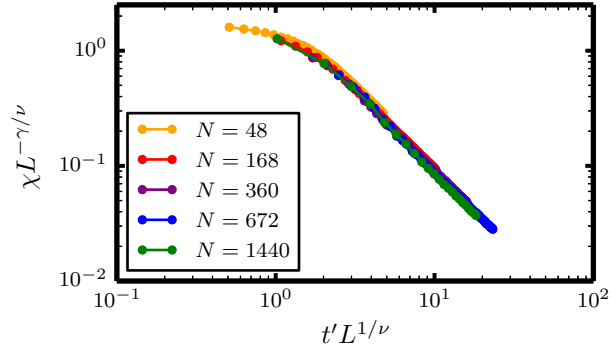

FIG. 20. Susceptibility for different system sizes at $c = 1/3$. Here, $t' = |1 - T_c/T|$. Averages were taken over $10^6$ MC sweeps.

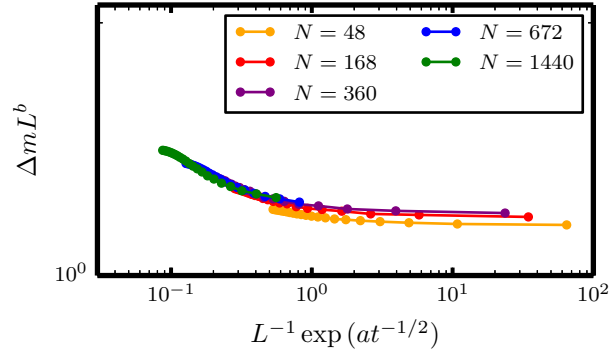

FIG. 21. Order parameter for different system sizes at $c = 1/2$. Here, $t = (T - T_c^{\text{lower}})/T_c^{\text{lower}}$. Data collapse is observed for all but the smallest system size. Averages were taken over $10^6$ MC sweeps.

where the functions $g_1$ and $g_2$ depend only on the quantity $L^{-1} \exp(at^{-1/2})$. Landau states that $a = 0.9$, $b = 0.095$, and $c = 1.72$. Using these values, we observed reasonable data collapse for $T_c^{\text{upper}} \approx 0.56$ and $T_c^{\text{lower}} \approx 0.46$. Plots of the scaled susceptibility and order parameter are shown in Figs. 21 and 22.

We confirmed that the correlation length upon approaching $T_c^{\text{upper}}$ scales as Eq. 128 by computing correlation

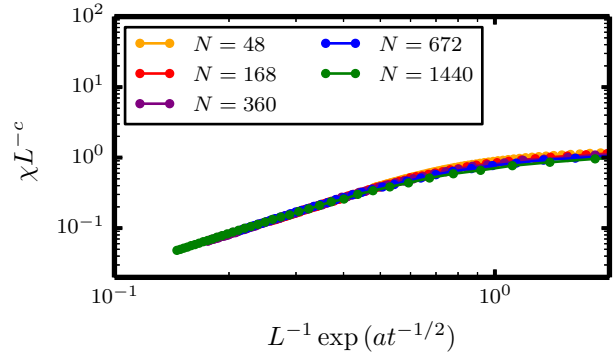

FIG. 22. Susceptibility for different system sizes at $c = 1/2$. Here, $t = (T - T_c^{\text{upper}})/T_c^{\text{upper}}$. Averages were taken over $10^6$ MC sweeps.

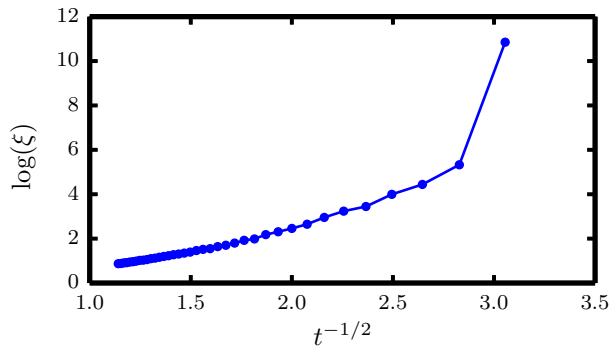

FIG. 23. Correlation lengths extracted from fit to $c(r)$. These correlation functions were obtained from MC simulations of an $N = 1440$ system at $c = 1/2$ using $10^7$ configurations for averaging. The linear behavior of the plot as $t \to 0$ indicates that the scaling of Eq. 128 is satisfied.

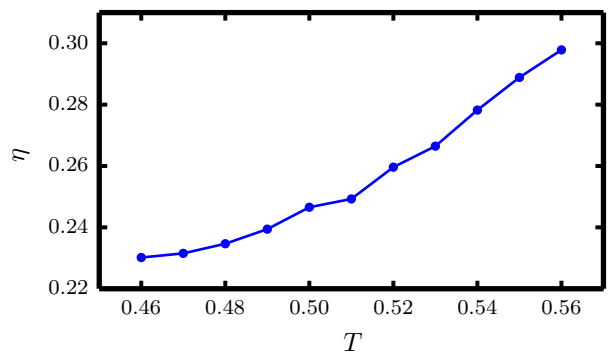

FIG. 24. Power law exponent $\eta$ extracted from fit to $c(r)$ between $T_c^{\text{lower}}$ and $T_c^{\text{upper}}$.

functions at $c = 1/2$ for various temperatures and fitting these to the form:

$$c(r) = \frac{Ae^{-r/\xi}}{r^\eta}. \tag{130}$$

Finite size effects cause an apparent divergence of $\xi$ prior to our independently measured value of $T_c^{\text{upper}}$. Nevertheless, the scaling of Eq. 128 appears to be satisfied, as shown in Fig. 23. Between $T_c^{\text{lower}}$ and $T_c^{\text{upper}}$, Landau states that while the correlation function should have power law behavior, the associated exponent will depend on temperature, i.e. $\eta = \eta(T)$. To confirm this, we fit correlation functions between $T_c^{\text{lower}}$ and $T_c^{\text{upper}}$ to a power law. A plot of the resulting exponent versus temperature is shown in Fig. 24, confirming that $\eta$ is indeed temperature-dependent.

### Coexistence Regions

To identify low-temperature coexistence regions (which are reasonably well-described by mean field theory combined with the quadratic construction,) we performed umbrella sampling simulations of our effective Hamiltonian at various temperatures. (Because the effective Hamiltonian contains only composition variables and not mechanical variables, it is easier to sample than the full Hamiltonian.) We then subtracted off the energy associated with coexistence at zero temperature, $E_{\text{coex}}(c)$. This simply consists of parabolic segments with curvature $-16$ connecting compositions $c = 0$ to $c = 1/3$, $c = 1/3$ to $c = 2/3$, and $c = 2/3$ to $c = 1$. These special compositions have energies $E(c = 0) = E(c = 1) = 0$ and $E(c = 1/3) = E(c = 2/3) = 8/3$. A plot of the free energies is shown in Fig. 25. There are clearly three distinct regions in this plot: $0 < c < 1/3$, $1/3 < c < 2/3$, and $2/3 < c < 1$, each of which, taken individually, develop non-convexities as temperature decreases. We associated these regions with coexistence. Specifically, we identified

coexistence lines as the boundaries of these non-convex regions, which we located with the standard double-tangent construction. We justify this procedure as follows. Recall our expression for the free energy of elastic coexistence:

$$f(c) = f(c_1) - \frac{\Delta c_1}{\Delta c_2 - \Delta c_1} \Delta f - Y \Delta l^2 \Delta c_1 \Delta c_2. \tag{131}$$

Non-convexity in the thermodynamic limit stems from the last term. We can remove this non-convexity by defining:

$$\tilde{f}(c) = f(c) + Y \Delta l^2 (c_1^* - c)(c_2^* - c), \tag{132}$$

where $c_1^*, c_2^*$ are the compositions of coexisting phases at zero temperature. We wish to prove that for $\tilde{f}(c)$, coexistence regions $[c_1, c_2]$ satisfy the conditions of the double tangent construction. To do so, we must show:

$$\left.\frac{\partial \tilde{f}}{\partial c}\right|_{c_1} = \left.\frac{\partial \tilde{f}}{\partial c}\right|_{c_2} \tag{133a}$$

$$\tilde{f}(c_2) - \tilde{f}(c_1) = \left.\frac{\partial \tilde{f}}{\partial c}\right|_{c_1} \Delta c. \tag{133b}$$

Eq. 133a is easy to demonstrate. The necessary derivative is:

$$\begin{aligned}\frac{\partial \tilde{f}}{\partial c} &= \frac{\partial f}{\partial c} + Y \Delta l^2 \left(2c - (c_1^* + c_2^*)\right) \\ &= \frac{\Delta f}{\Delta c} - Y \Delta l^2 \left[(c_1^* + c_2^*) - (c_1 + c_2)\right],\end{aligned} \tag{134}$$

which is independent of $c$, proving the equality. Eq. 133b is also straightforward to prove. The left hand side can be written:

$$\begin{aligned}\tilde{f}(c_2) - \tilde{f}(c_1) &= \Delta f + Y \Delta l^2 \left[(c_1^* - c_2)(c_2^* - c_2) - (c_1^* - c_1)(c_2^* - c_1)\right] \\ &= \Delta f + Y \Delta l^2 \left(-c_2 c_2^* - c_1^* c_2 + c_2^2 + c_1 c_2^* + c_1^* c_1 - c_1^2\right),\end{aligned}$$

as can the right hand side:

$$\begin{aligned}\left.\frac{\partial \tilde{f}}{\partial c}\right|_{c_1} \Delta c &= \Delta f - Y \Delta l^2 \left[(c_1^* + c_2^*) - (c_1 + c_2)\right] (c_2 - c_1) \\ &= \Delta f + Y \Delta l^2 \left[-(c_1^* + c_2^*)(c_2 - c_1) + (c_1 + c_2)(c_2 - c_1)\right] \\ &= \Delta f + Y \Delta l^2 \left(-c_1^* c_2 - c_2 c_2^* + c_1^* c_1 + c_1 c_2^* + c_2^2 - c_1^2\right).\end{aligned}$$

The right and left hand sides are clearly equal, showing that the conditions of the double tangent construction are indeed satisfied. For any finite-sized system, the free energy will not be strictly convex – surface tension will result in a non-convex contribution to $f(c)$ scaling as $O(N^{-1/2})$ (in two dimensions.) Knowing that in the thermodynamic limit this contribution vanishes and the free energy of coexistence satisfies the above conditions, we applied the double tangent construction to the finite system free energy and took the coexisting compositions thus inferred as proxies for the coexisting compositions for an infinite system.

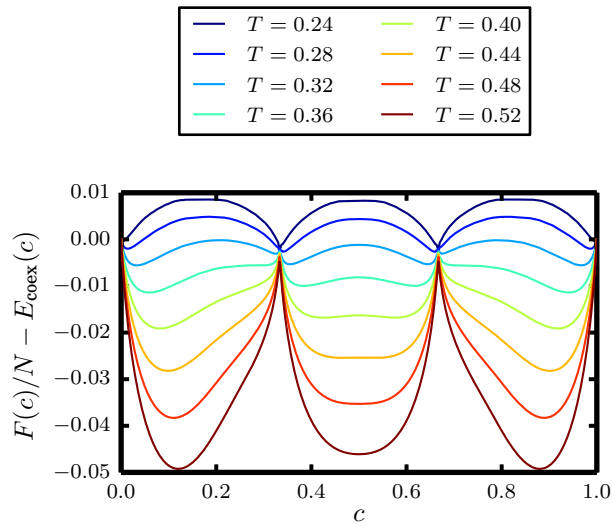

FIG. 25. Free energy minus the coexistence energy for different temperatures. Free energies were obtained for systems with $N = 360$ atoms, using umbrella sampling with 100 evenly-spaced harmonic windows. In each window, we performed $10^7$ MC sweeps of equilibration and subsequently collected data every sweep for $10^7$ MC sweeps.


\fi

\end{document}